\documentclass{article}

\PassOptionsToPackage{numbers, compress}{natbib}

    \usepackage[preprint]{neurips_2022}

\usepackage{svg}
\usepackage{pdfpages}
\usepackage{adjustbox}
\usepackage[utf8]{inputenc} 
\usepackage[T1]{fontenc}    
\usepackage{hyperref}       
\usepackage{url}            
\usepackage{booktabs}       
\usepackage{amsfonts}       
\usepackage{nicefrac}       
\usepackage{microtype}      
\usepackage{xcolor}         
\title{The Gradient of Generative AI Release: \\Methods and Considerations}

\author{%
  Irene Solaiman\\
  Hugging Face\\
  \texttt{irene@huggingface.co} \\
}

\begin{document}

\maketitle

\begin{abstract}
    As increasingly powerful generative AI systems are developed, the release method greatly varies. We propose a framework to assess six levels of access to generative AI systems: fully closed; gradual or staged access; hosted access; cloud-based or API access; downloadable access; and fully open. Each level, from fully closed to fully open, can be viewed as an option along a gradient. We outline key considerations across this gradient: release methods come with tradeoffs, especially around the tension between concentrating power and mitigating risks. Diverse and multidisciplinary perspectives are needed to examine and mitigate risk in generative AI systems from conception to deployment. We show trends in generative system release over time, noting closedness among large companies for powerful systems and openness among organizations founded on principles of openness. We also enumerate safety controls and guardrails for generative systems and necessary investments to improve future releases.
\end{abstract}

\section{Introduction}  \label{intro}
What constitutes a robustly safe and responsible release of new AI systems, from components such as training datasets to model access itself, urgently requires multidisciplinary guidance. There is no overarching standard or standards-making body to form consensus for what constitutes responsible release. This is particularly true for generative AI systems, which can be leveraged for a broad range of tasks and are sometimes referred to as “general-purpose” \cite{Trajtenberg18}. A sub-type of foundation models \cite{DBLP:journals/corr/abs-2108-07258}, these systems generate outputs across modalities such as text and image. They can be applied to both beneficial and harmful tasks. For example, language models can be adapted to tasks such as grammar correction and translation, or be used for phishing and spam. The relative novelty means new uses, and misuses, of these systems are still being discovered. Carefully considering release strategies for present and less powerful systems better prepares and sets precedent for the AI community and the broader affected community as increasingly powerful systems are deployed. Due to the fast pace of AI progress and release, developers, deployers, researchers, and policymakers must take action via community discussions, guardrails, and investments.

The current state of generative AI system release is largely determined by the organizations developing systems. Developers are most likely to best understand the system, but understanding impact and ripple effects requires multidisciplinary expertise that is rarely housed in one organization. Waiting for long-term evidence of consequences is infeasible for high-risk and powerful systems. A strictly closed and vertical process to commercialization can lead to concentrated power among high-resourced organizations. An open process without ethical considerations can inflict and exacerbate risk and harm, from misuse to bias.

Many components make up a system throughout its lifecycle, from training data to computing power. This paper will primarily focus on the cumulative release of a model and its components by outlining key considerations in release; release options along the gradient; the timeline of released systems; and necessary investments to improve safe releases. 

\section{Previous Work}
Discussion about safe release has been ongoing in the AI research community, but there is no standards body or default convener for these discussions. Initiatives such as the Partnership on AI’s Publication Norms for Responsible AI \cite{staff_2022} and Stanford University’s Call for Community Norms for the Release of Foundation Models \cite{liang_bommasani_creel_reich} have made space for discussion around the many options for system releases \cite{sastry}, and the many components involved and available platforms that have made options complex. 

At the core of release considerations is the tension between openness that shares rather than concentrates power, and closedness that minimizes potential harm and risk. Broadly, safety and risk control development lags behind system development; for example, tools for detecting generated outputs underperform as systems become increasingly powerful \cite{DBLP:journals/corr/abs-2002-03438}. For researchers in natural language processing, whether developers should be ethically responsible for downstream misuse of a publicly released system is contentious, with about half of researchers believing professionals should be responsible\cite{https://doi.org/10.48550/arxiv.2208.12852}.

Parallel fields such as open-source software deployment can share informative lessons, such as the ability for open source software communities to enable community research and crowdsourcing work such as discovering vulnerabilities \cite{10.1145/2145204.2145396}. Examining the specific use case of open-source Deepfakes highlights the difficulty of managing downstream harms in real-time and the risk of safety controls being seen as futile \cite{10.1145/3531146.3533779}. While generative AI research would also greatly benefit from this community insight, lessons from software are not often directly applied due to substantial difference in functionality \cite{DBLP:journals/corr/abs-2001-00463}. Ultimately personal values around openness are a large factor in decision making \cite{DBLP:journals/corr/abs-1910-01170}, and tensions can be further examined by output modality \cite{bergman-etal-2022-guiding}. 

\section{What is Being Released?}
The parts of an AI system considered in a release can be broken into three broad and overlapping categories: 
\begin{itemize}
  \item access to the model itself, 
  \item components that enable further risk analysis, 
  \item and components that enable model replication. 
\end{itemize}
Components are organized based on their most straightforward use. There is overlap among these components; the same model cannot be replicated without its component for risk analysis, such as its entire original training data, even if all replication components are available. Conversely, components for replication can also be analyzed for social impacts such as biases. 

\subsection{The Model Itself}
Access to the model itself includes the model weights and the ability to query, adapt, or otherwise examine and conduct further research into a model. The range of access is expanded according to the gradient in Figure \ref{fig:gradient}. 

\subsection{Components for Risk Analysis}
These components are the parts of system development that could provide further insight into the model; the model's capabilities; the decision making process on what data was collected and how; and documentation of the process. Additionally this details system risks, training data, fine-tuning data, and information on people and human crowdworkers involved in adapting the model through methods such as reinforcement learning with human feedback. This also includes evaluation results: published results from any evaluations that researcher and developers may have run on the base model. These components may be withheld due to intellectual property (IP) rights, consent, or privacy concerns.

\subsection{Components for Replication}
These components include a technical paper detailing the model training process and code used to train the model, as these can ease replication efforts. This also includes training information such as configuration settings (e.g. batch size), and telemetry collected during training (e.g. training loss). These components may be withheld for competition, IP, and misuse reasons. They are also high risk for misuse concerns as they can be repurposed or adapted to malicious or otherwise harmful use cases \cite{vee_2022}. 

\section{Key Considerations in Release}
Deployers should weigh the following considerations when making release decisions. Risks and threats from increasingly powerful systems are difficult to enumerate and assess, especially since malicious actors and their incentives are constantly evolving \cite{10.1145/3372823}. Taxonomies of ethics and risks of specific systems \cite{https://doi.org/10.48550/arxiv.2112.04359} can serve as a framework for potential harms. Specific considerations across all generative systems are listed below. 

\subsection{Concentration of Power}
One of the most prominent arguments for providing access to systems is to avoid concentrating the level of power that high-resource organizations are collecting as one of the few groups capable of developing and deploying these systems. Large technology companies are able to create powerful AI systems because of their access to training data, computing infrastructure, and commercial capabilities for deploying that system. This monopolization also gives these high-resource institutions more influence in AI development, the behavior of these systems, and the narrative and direction of the field \cite{benkler_2019}. Although these companies may provide access or even open-source their systems, contributions to system development are limited to people and resources working towards that company’s interests \cite{10.1145/3488666}. Large companies are often geographically concentrated in Western countries whereas systems are deployed globally, which can asymmetrically impose cultural values \cite{DBLP:journals/corr/abs-2007-04068}. These companies can also punish pushback or dissent \cite{dave_dastin_2020}. The people most affected and exploited by AI systems are rarely found in large technology companies. They must be empowered to shape systems that also benefit them, or to opt out of interaction with AI entirely \cite{Kalluri_2020}.

\subsection{Exacerbating Disparate Performance and Harmful Social Impacts}
The fewer perspectives that are incorporated into the system development process leads to higher likelihood the system performs disparately for different groups. AI systems can propagate harms such as exacerbating social inequity \cite{noble_2018, benjamin_2020, hovy-spruit-2016-social} and harmful biases \cite{doi:10.1126/science.aal4230, blodgett-etal-2020-language}, which can be further amplified in larger systems as scale increases \cite{10.1145/3442188.3445922}. Means of measuring and mitigating risk in these systems are largely cultural and context-dependent \cite{talat-etal-2022-reap}. The many technical and social aspects of AI systems \cite{DBLP:journals/corr/abs-2108-07258} require robust research \cite{DBLP:journals/corr/abs-2111-15366} conducted with communities affected \cite{10.1145/3531146.3533083} to ensure these systems benefit and do not exploit marginalized groups, if the systems are to be deployed among these groups.

\subsection{Malicious Use and Unintentional Misuse}
With more modalities of AI generation improving in output quality, from high quality text to high quality images, the potential for harmful use cases also increases. Malicious uses such as the creation of deepfake imagery \cite{DBLP:journals/corr/abs-1909-11573}, AI-generated disinformation \cite{https://doi.org/10.48550/arxiv.2301.04246}, and illegal and disturbing material \cite{simonite_2021}, can cause severe emotional harm at the individual level and destructive institutional harm at the societal level. Furthermore, malicious actors \cite{DBLP:journals/corr/abs-1802-07228} have historically worked to circumvent safety controls. Threat modeling will necessarily differ by modality, but as systems improve in types of outputs such as code generation, potential harms can also broaden \cite{https://doi.org/10.48550/arxiv.2207.14157}. While limiting access can prevent some malicious uses and is often a suggested action to minimize misuse \cite{DBLP:journals/corr/abs-1907-11274}, systems can still be vulnerable to attacks with only querying functionality available \cite{DBLP:journals/corr/abs-2012-07805}. 

\subsection{Auditability}
The question of auditability addresses who is conducting audits and the level of access required to effectively examine an AI system. Auditing must be considered both pre- and post-deployment as impacts from a system may not be detectable pre-deployment and when deployed, impacts may be difficult to trace back to a specific system \cite{10.1145/3351095.3372873}. The actors conducting and capable of conducting audits will likely require some level of technical skill even when numerous no-code tools are built. The size of the system and its components also determine auditability; the datasets that large generative AI systems are trained on are not only difficult to analyze at scale, but few tools exist to analyze large static datasets \cite{Cai2015TheCO}. Formal audits alone cannot be the only insight or governance of a system \cite{10.1145/3514094.3534181}. 

\subsection{Accountability in Case of Harm}
In the case that an AI system harms or is connected to harming people, who or what is to be held accountable is unclear. More open and deployed systems have a higher likelihood of a broader reach and therefore a higher chance of harm. Since harm is not explicitly defined and not always physical, what constitutes harm can have a large range. The range may include encouraging physical harm, propagating social harms such as identity stereotypes, and more abstract harms such as lack of access to a system lowering opportunities for a specific group. Work to characterize sociotechnical harms can narrow the scope \cite{https://doi.org/10.48550/arxiv.2210.05791}.

\subsection{Value judgments for gating and limiting access}
A base generative AI system is capable of many types of content, making content moderation complex \cite{gillespie_2021}. What constitutes appropriate outputs is influenced by religion, cultural, and personal beliefs. What content can and should be limited, filtered, and gated is also vague. For example, sexual content may not be inherently unsafe to generate in some cultures, but may be subject to local laws. Technical filters may not be able to distinguish between sexual content and nonsexual nudity, and may not be able to distinguish between consensual and non-consensual content. While most difficult without specific use cases or context, specific applications face the same challenges.

\section{The Gradient of System Access}
Once considerations are taken into account, the group determining release method must choose if the system and its components are publicly acknowledged and released. The below gradient of release options are based on five years (2018 - 2022) of publicized generative AI systems. This gradient of options serves as a framework and does not fully capture the nuance of the many components and details in a system release.

\begin{figure}[h!]
  \centering
  \includegraphics[width=\textwidth]{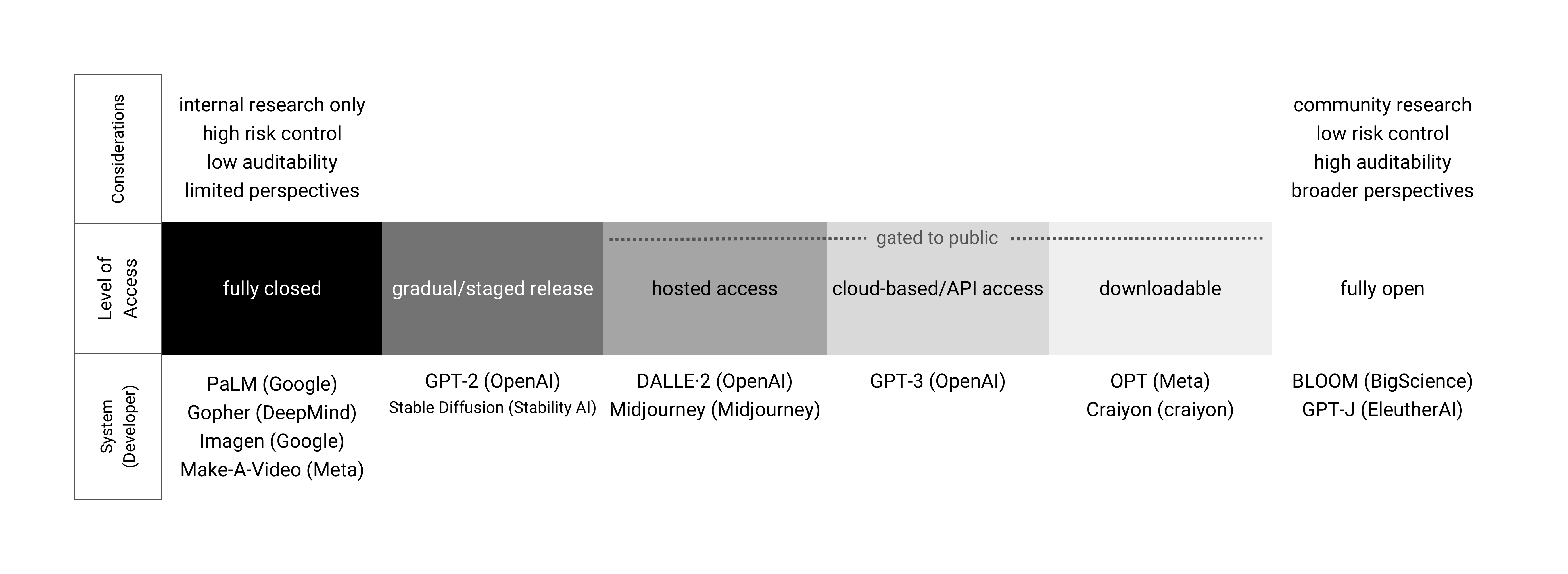}
  \caption{Considerations and Systems Along the Gradient of System Access}
  \label{fig:gradient}
\end{figure}

Figure \ref{fig:gradient} shows the tradeoffs in considerations along the gradient; as systems become more open they better enable audits and community research but are more difficult to control for risks. 

Below the gradient are examples of generative systems placed according to their original release method upon announcement; for example, GPT-2 may fall under “Downloadable” today, but was originally released as “Gradual/Staged”. 

\subsection{Fully Closed} 
When all aspects and components of a system are inaccessible outside the developer organization, or even closed outside a specific subsection of an organization, the system is fully closed. At the furthest end of the spectrum, the system’s existence is unknown outside a select group within the developer organization, even after full training. A fully closed system may or may not include some form of public announcement that the system exists. These systems can only be researched by the developer organization, which is often a high-resource organization such as the Alphabet companies Google and DeepMind. Some publicly-known systems, such as Google’s Imagen \cite{https://doi.org/10.48550/arxiv.2205.11487} and DeepMind’s Gopher \cite{DBLP:journals/corr/abs-2112-11446}, are examples. Public engagement may come from the system being deployed in a commercial application or the public calling out biases and notable social aspects of a system from public releases. These releases can be cherry-picked, for example showing only non-human animals or human silhouettes \cite{https://doi.org/10.48550/arxiv.2209.14792}, which does not give robust insight to broad capabilities or social impacts such as biases. 

\subsection{Gradual/Staged Release}
This method refers to releasing a system in stages or gradually over a predetermined amount of time. The time between stages is intended for investments that minimize risk such as monitoring for malicious actor activity and conducting research on potential harms. In 2019, OpenAI stage-released language model GPT-2 in four sizes by increasing parameter count over nine months while conducting research internally and with external partners \cite{DBLP:journals/corr/abs-1908-09203}. This sparked debate among some \cite{lipton_2019}, but still is a recommended tactic among others \cite{https://doi.org/10.48550/arxiv.2210.04610}. In 2022, Stability AI’s Stable Diffusion \cite{mostaque_2022} initially approached a stage release by providing access to a hosted model before releasing the model weights. However, model weights were leaked 12 days after their initial hosted release. This exemplifies the need to inject safety protocols and prevent leakage during this approach. While there is no standardized time frame for staged releases, generally substantial sociotechnical research requires multiple weeks, months, and sometimes years.

\subsection{Gated to Public Access (Including Paid and Free)}
When providing access to a system without fully opening all components, actors deciding release method may choose to place access limitations. Above the infrastructural limitation options, namely hosting, cloud-based access, or fully downloadable access, is the choice to make the release gated or public. Gating system access is a selective process by group of people used to block high risk or out-of-scope use cases. Limited access can make enforcing controls easier; for example, system deployers withhold the right to revoke access in a gated and hosted access setting. However, gating downloadable systems is unreliable as a technical mechanism; the network effect of researchers sharing within the same circles \cite{liang_bommasani_creel_reich} can provide a loophole to gating. The releasing organization cannot fully monitor whether users are sharing access through screen-sharing, credential-sharing, or simply sending components such as model weights to unauthorized users. This does not mean this is an ineffective guardrail, as it still creates barriers to sharing the model. The deploying organization will still be making critical decisions that contribute to concentration of power.

\subsubsection{Hosted Access}
System deployers may provide access to the model itself by hosting the model on their own servers and allowing surface-level interfacing. Access can differ depending on the interface's usability, especially for users with minimal or no experience with these systems. Generally, users are unable to perform tasks outside what is prescribed: usually simple input-output probing. This method is specific only to model access, not access to other system components. Examples include Midjourney, which allows users to interact with its image generation model, via Discord bot or web interface \cite{midjourney_2022}. When optimized for usability and dialogue, as seen with OpenAI's ChatGPT \cite{openai_chat}, broader perspectives can interact with and the model, but raises misuse concern and ethical challenges \cite{reich_2022}. This method can also transition to API or downloadable, as seen with OpenAI’s DALL$\cdot$E 2 \cite{https://doi.org/10.48550/arxiv.2204.06125} was switching to API-based access seven months later \cite{openai_2022}. While this method provides some model access, it limits external research ability.

 \subsubsection{Cloud-based/API Access}
Cloud-based access or access provided via application programming interface (API), provides more insight and researchability into a model than Hosting, but still allows for restrictive functionality. Some APIs only allow for querying, such as OpenAI’s original GPT-3 release via API \cite{openai_2020}. Additional functionality can be added, such as fine-tuning via API. Similar to Hosting, this method is specific only to model access. Non-released components and system information can be determined via tools such as EleutherAI’s evaluation harness, used to determine GPT-3 parameter sizes via OpenAI’s API \cite{gao_2021}. This method is favorable for structured access where research is possible but can still be tracked and is unlikely to create a “modified version” \cite{https://doi.org/10.48550/arxiv.2201.05159}. Cloud-based access can track users and their activity to monitor for risky behavior. This can also better enforce safety controls such as rate limiting. . 

\subsubsection{Downloadable}
The main distinction between downloadable and fully open systems is the withholding of system components, such as training dataset availability. Downloadable systems can also be gated. Downloadability does not inherently imply full access to any user granted access, as the size of a model can limit who is capable of running a modal locally. Personal and standard consumer hardware is unlikely to support large and powerful models. The infrastructure needed to run large models creates an access barrier. In response, industry \cite{microsoft_2022} and public initiatives \cite{nairrtf_2021} are creating accessible infrastructure for researchers. Downloadable models better enable robust research, but are difficult to track for potential misuse or harm. This method also eases user ability to erode or disable safety controls, such as content filters.

\subsection{Fully Open}
When all aspects of the system are accessible and downloadable, including all components, the system is fully open. These systems cannot be gated and by definition are fully public. For the purposes of this framework, a basic level of accessibility and documentation across components qualifies a system release as “fully open”, but releases may differ in documentation detail and levels of granularity. The most prominent fully open systems were developed by organizations founded on the principle of openness. EleutherAI is a decentralized collective also prioritizes transparency and has released all system components, as seen in their GPT-J \cite{mesh-transformer-jax} and GPT-Neo \cite{gpt-neox-20b} language models and the Pile dataset \cite{DBLP:journals/corr/abs-2101-00027}. The BigScience global research community of over 1000 researchers developed the BLOOM language model in the open \cite{https://doi.org/10.48550/arxiv.2211.05100}. Over 30 working groups covered aspects from dataset creation to carbon footprint to modeling approach to optimize for a multilingual system created transparently \cite{bigscience-2022-bigscience}. While openness does enable broader research that can engage many peoples, it can also enable dangerous uses and model creation \cite{vee_2022} and controls can be difficult to enforce. 

\section{Trends in System Releases}
We analyze release trends across prominent base generative AI systems; this does not include fine-tuned or updated systems such as models that undergo reinforcement learning with human feedback. These figures are based on tracking and evaluation initiatives \cite{LiaoModelTracker2022, https://doi.org/10.48550/arxiv.2211.09110, talat-etal-2022-reap}, are not exhaustive, and intend to show release trends over time. 

\subsection{Timelines for Large Language Models}
\begin{figure}[h]
  \centering
  \includegraphics[width=\linewidth]{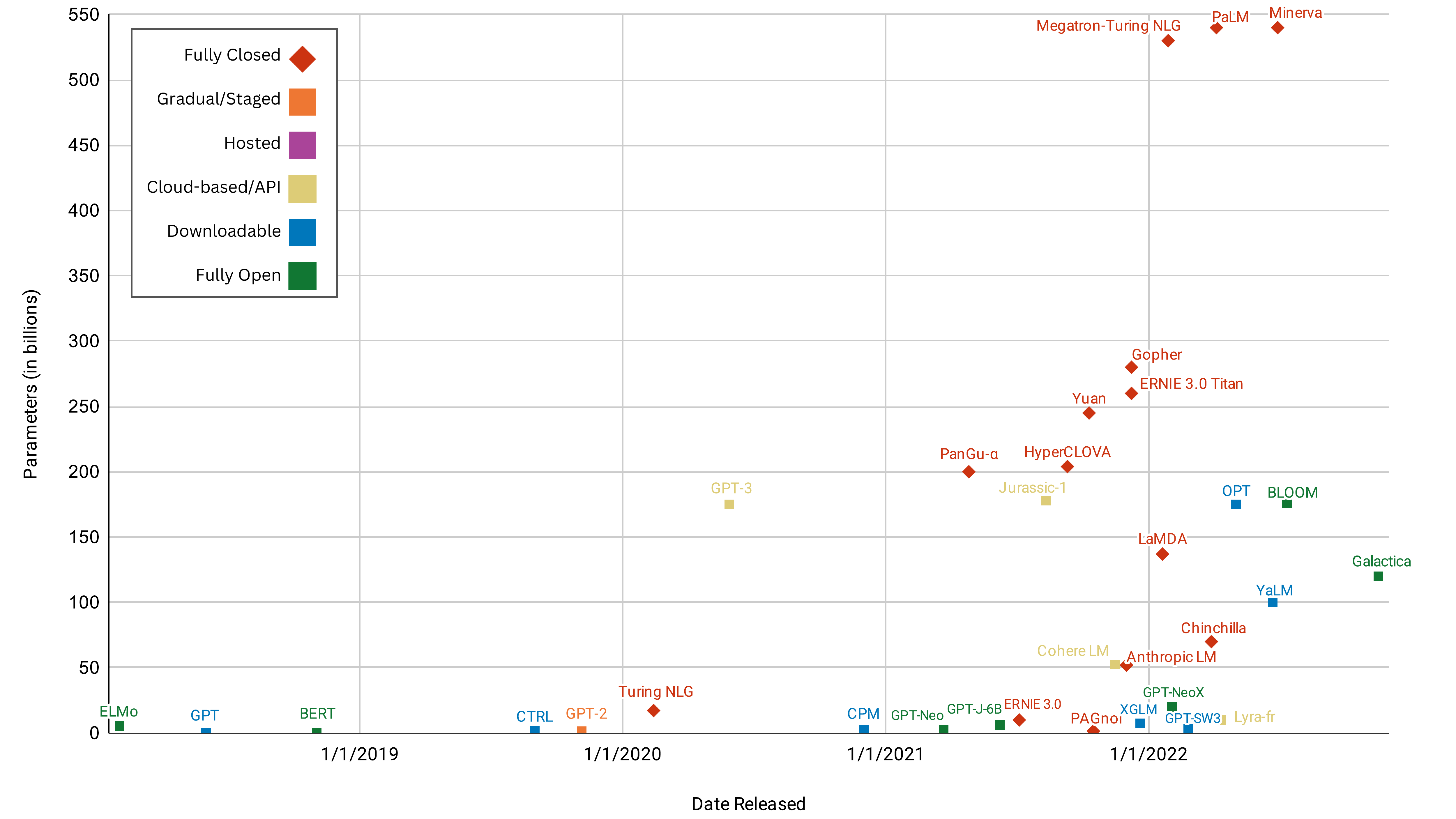}
  \caption{Language Model Release Method By Parameter Count Over Time}
  \label{fig:lm_release}
\end{figure}

When examining systems by the original method of release over time, trends seen in Figure \ref{fig:lm_release} show closing and limiting language model access as more common since GPT-2’s staged release. Language models with fewer than six billion parameters have generally been towards the open end of the gradient, but more powerful models, especially from large companies, tend to be closed. This can be due to their requiring deeper consideration and safeguards due to risk potential, but Figure \ref{fig:lm_release} also illustrates the high number of large companies able to develop and close language models.

\pagebreak
\subsection{Timelines for All Modalities}
As more generative modalities are developed, from image to audio to video, they face similar release decision challenges. 

\begin{figure}[t]
  \centering
  \includegraphics[width=.92\linewidth]{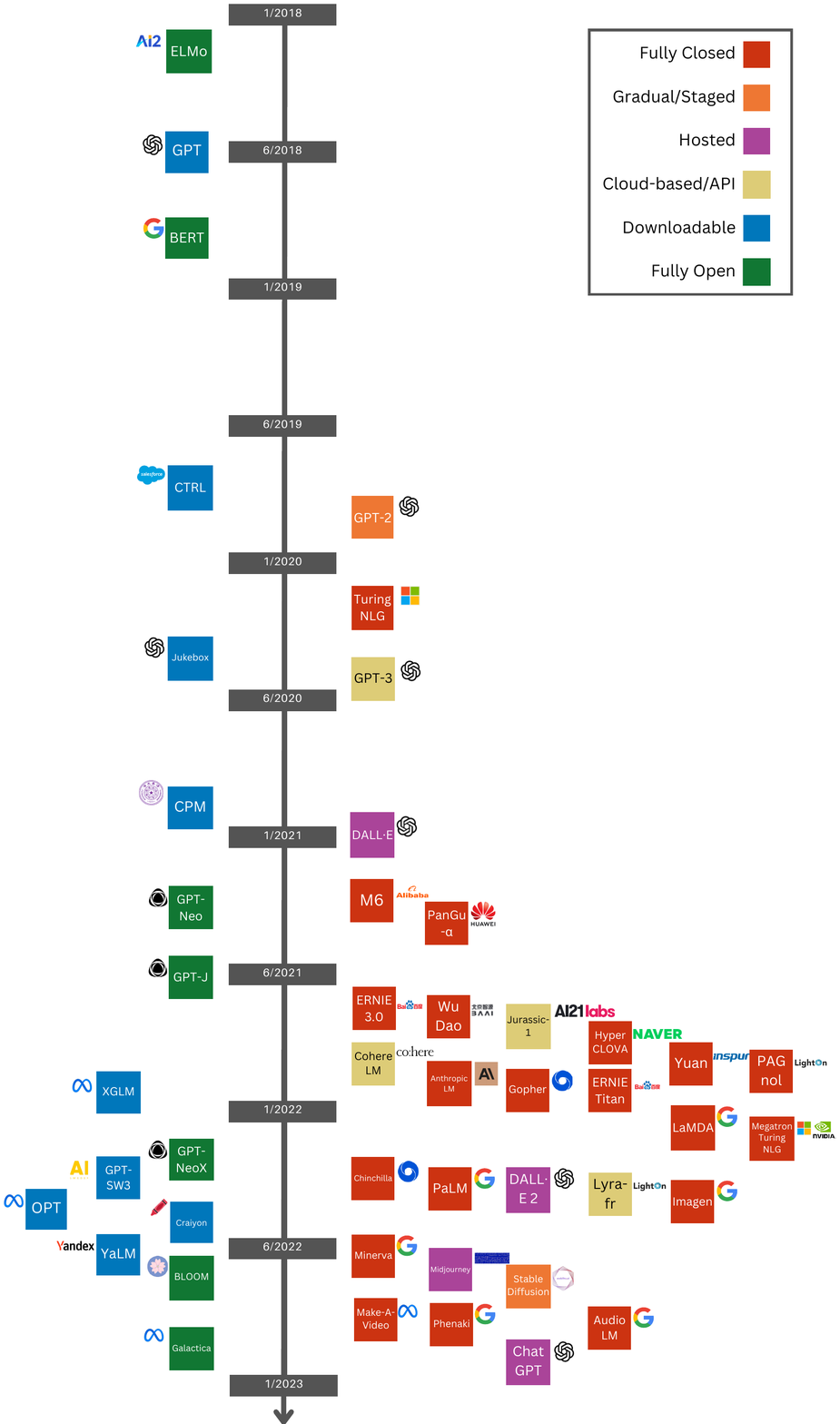}
  \caption{Release Methods Over Time (All Modalities)}
  \label{fig:all_modalities}
\end{figure}

 Figure \ref{fig:all_modalities} shows system release over this same time period. As there is no standard means to compare capabilities across modalities, all levels of system capability are placed equally. Again, trends show openness until GPT-2's staged released. This timeline also shows a sharp increase in the amount of systems developed, and closed, after 2021. The systems most commonly toward the open end of the gradient are developed by smaller organizations founded with the intent to be open. 

 Conversely, many systems from large companies\footnote{See Appendix \ref{Appendix:A} for logo and developer key} are becoming closed or have closed components. OpenAI is the most common company to restrict but not fully close or open access. Alphabet companies Google and DeepMind are most common among closed systems. Across modalities, large companies have steered toward closedness. Open initiatives from large companies are shown to release a downloadable model trained on public datasets crafted by other organizations, as seen with Meta’s OPT-175B \cite{https://doi.org/10.48550/arxiv.2205.01068}. It is unclear at this time whether movements towards openness will pressure historically closed organizations to adjust their release strategies.

\section{Safety Controls and Guardrails}
A combination of controls and guardrails, largely from the developer and deploying organizations but also from external researchers, can complement each other in order to address the above considerations and risks. Many of these methods are pioneered and honed in research environments and outside developer organizations. Individually, no one control can serve as a panacea. While it is possible to add controls and guardrails long after deployment, these options are most effective when deployed simultaneously with system release. 

\subsection{Documentation and Transparency}
Structured documentation that clearly communicates critical information about each component of the system gives further insight to the system and can take many forms. Proposed approaches to documentation at dataset and model levels have proven successful; without any enforcement mechanisms, many releases across AI companies include some form of this documentation. \textit{Datasheets for datasets} \cite{DBLP:journals/corr/abs-1803-09010} communicates aspects of datasets such as creators’ motivations, collection process, and overall composition. Meta’s OPT-175B release included a datasheet in its appendix \cite{https://doi.org/10.48550/arxiv.2205.01068}. \textit{Data statements for natural language processing} \cite{bender-friedman-2018-data} are another popular tool more tailored to language-based systems, seen used by bias measurement dataset CrowS-Pairs \cite{DBLP:journals/corr/abs-2010-00133}. \textit{Model cards} \cite{DBLP:journals/corr/abs-1810-03993} have been popular as seen in Google’s PaLM \cite{https://doi.org/10.48550/arxiv.2204.02311}, OpenAI’s GPT-2 \cite{openai_2019} and GPT-3 \cite{openai_gpt3}, and Runway Research and Stability AI’s Stable Diffusion \cite{Rombach_2022_CVPR}. Model cards are deployed across Hugging Face’s platform and have evolved to be interactive \cite{Crisan_2022}. \textit{System cards} \cite{Procope_2022} blend datasheets and model cards and have been used for DALL$\cdot$E 2 \cite{mishkin2022risks}.

\subsection{Technical Tools}
Technical tools can address specific technical safety concerns, but cannot be a substitute for addressing complex societal problems. In some cases, technical tools can create new social harms and should therefore be vetted and combined with other guardrails.

\subsubsection{Rate Limiting}
Constricting the amount of outputs a user can generate via cloud-based access is a popular means of preventing attacks and harmful generations. Rate limiting also helps a system perform well and protect underlying infrastructure from being overloaded. This defensive measure can be enforced with common strategies such as a token bucket, which tracks and limits usage according to a set number of tokens that can refresh or accumulate on a predetermined time frame. As an example, OpenAI’s DALL$\cdot$E 2’s public API rate limits external users \cite{openai_rate}. This can be adapted for users whose applications have been cleared as safe.

\subsubsection{Safety and Content Filters}
Filters developed to trigger blank responses when given an unsafe input are popularly deployed across varying levels of access. This can help block illegal and egregious content. Developers selecting these trigger categories must make normative judgements about what input content blocks generation. Stable Diffusion’s safety filter was found to primarily prevent generations with sexually explicit content but not violence and gore \cite{https://doi.org/10.48550/arxiv.2210.04610}, which is a normative judgment about the safety of both categories in generated images. Blocking generations for socially sensitive topics can result in entire identity groups being blocked. Lessons from social media platform content moderation highlight harms such as community erasure, especially among marginalized groups \cite{10.1145/3479610}.

\subsubsection{Detection Models}
While methods to detect AI generated outputs can vary and include human detection, detection models can be a helpful tool, especially for less powerful generative systems. While the human eye alone can detect outputs from less powerful systems, such as Craiyon for AI-generated images, detection models can have higher accuracy for more powerful systems. This is particularly important when models are deployed in high-stakes settings \cite{kreps_mccain_brundage_2022}. As system output quality improves, that distinction becomes more difficult for both humans and AI detection models. Approaches to detection can be tailored to modality \cite{DBLP:journals/corr/abs-2011-01314} such as text \cite{gehrmann-etal-2019-gltr}, and include human annotation \cite{DBLP:journals/corr/abs-2107-01294}. Detection models can also differ based on type of generation within a modality, such as facial generations \cite{Gragnaniello2022}.

\subsubsection{Hardcoding Responses}
Predetermined safe outputs triggered for a given input can be hardcoded into a model interface. This can aid legal compliance or provide standardized responses for high-risk inputs. Similar to filtering, determining trigger inputs or trigger categories requires normative judgements about what constitutes unsafe inputs and what constitutes appropriate outputs. This can not only lead to community erasure, but also impose these normative beliefs onto users.

\subsubsection{Watermarking}
The concept of digital watermarking media can be transferred to AI systems to protect against model theft, protect IP, and more easily identify AI-generated outputs. Encoding a unique identifier in generated outputs can aid in detecting media as AI-generated and synthetic and trace the output to a specific model. Research strives ensure these watermarks are invisible to the human-eye \cite{DBLP:journals/corr/abs-1909-01285}, do not affect output quality \cite{https://doi.org/10.48550/arxiv.2301.10226}, and tamper-proof from model attacks and alterations like fine-tuning via methods such as embedding noise as watermarks \cite{10.1145/3196494.3196550}. Different approaches to watermarking can be deployed for different needs, from the embedding method to easily determine whether an output is synthetic to linking watermarks to a model owner’s identity for authentication purposes \cite{10.3389/fdata.2021.729663}. There are no current prominent successful case studies as watermarking has not yet been publicly deployed at scale for large generative systems.


\subsubsection{Model Weight Encryption}
Encryption can be used in order to protect model weights, often to protect from model stealing and to protect IP. This allows only an authorized user with the key to use the model. \cite{10.1145/3505634}’s proposed NN-Lock does not change model structure so as not to adversely affect model performance. \cite{10.5555/3437539.3437711} proposed an obfuscation framework that only authorizes users with a trustworthy hardware device. \cite{DBLP:journals/corr/abs-2011-13564} notes many existing IP protection methods are not robust to model attacks and are not suitable for commercial purposes as they verify model ownership but not user identities.

\subsubsection{Updating, Adapting, or Retraining models}
Models can be adapted in a way that mitigates risk. Popular methods include fine-tuning; for example fine-tuning GPT-3 on values-targeted datasets \cite{solaiman2021process}, or fine-tuning LaMDA on annotated data to improve factual grounding \cite{https://doi.org/10.48550/arxiv.2201.08239}. Another method is reinforcement learning with human feedback as seen with InstructGPT \cite{https://doi.org/10.48550/arxiv.2203.02155} and its open-source replication effort at CarperAI \cite{carperai_2022}. These methods result in new models different from their base models, but often improved along a safety parameter.

\subsection{Community and Platform Efforts}
Community-driven approaches to risk mitigation leverage new and varied viewpoints. Bounty programs, from bug bounties to bias bounties \cite{rubinovitz_2018}, can raise unforeseen safety issues and strengthen trust in a system \cite{https://doi.org/10.48550/arxiv.2004.07213}. \textit{Bias bounties} by nature benefit from diverse perspectives. \textit{Community moderation}, \textit{community-based content flagging}, and \textit{naming and shaming techniques} on a platform enables users to determine and stop harmful content before it escalates. \textit{Monitoring and logging inputs} by a user on the backend helps track trends in harmful or extremist behavior.

\subsection{Organizational and Platform Policies}
Organizational and platform policies can guide and enforce safe human interaction with generative AI systems. These policies can have drawbacks; they may protect from harm, but also limit beneficial uses. For example, limiting access in a region under active war can prevent disinformation generation but also general access. \textit{Internal risk policies} should provide a process for what considerations must be weighed and how to evaluate each prior to determining release options. If the system is deployed on a platform or on a given interface, \textit{a code of conduct} for engaging with the platform and other users prevents direct harm on the platform at risk of losing platform access. \textit{Mandating user accounts} on a platform helps track specific users and their activity, which supports community and platform efforts. \textit{Sharing policies} that outline what can and cannot be posted on other platforms or for uses outside of personal use prevents harmful content from spreading and inciting further harmful content.

\subsection{Legal Recourse}
Legal measures such as licenses are an enforceable control when a user uses a system in a way the deployer prohibited. The Responsible AI License (RAIL) places behavioral use conditions on a model, with the model owner owning the license and responsibility for pursuing enforcement if need be \cite{DBLP:journals/corr/abs-2011-03116}. Both BigScience’s BLOOM model \cite{bigscience_2022} and Runway Research and Stability AI’s Stable Diffusion \cite{rombach_esser_2022} use RAILs. Licenses are difficult to enforce for downloadable or fully open systems, as model behavior and uses cannot be fully monitored. Legal enforcement can also be costly in terms of both time and financial resources. Example cases studies are examined in \cite{DBLP:journals/corr/abs-2011-03116}.

\section{Necessary Investments for Responsible Release}
Developers and researchers must listen and leverage multidisciplinary and often external expertise, especially for guardrails. Policymakers must mandate safety where possible and technically feasible, and provide resources for the under-resourced. Regardless of level of access, generative AI systems are capturing, reflecting, and amplifying aspects of society that require multiple perspectives in exploratory and risk control research. Since a system cannot be fully safe or unbiased for all groups of peoples and there is no clear standard for when a system is safe for broad public release, further discourse across all affected parties is needed. Research and decisions made now will inform considerations for increasingly powerful systems across modalities in the future, making early investment crucial.

\subsection{Accessible Interfaces and Low and No-code Tools}
In order to make generative systems accessible to the many peoples they affect, means of interacting with a system, such as a model demo, are needed. A clean, easily usable interface that accommodates disabilities and all levels of technical comfort significantly improves accessibility. This step towards further openness can push a system toward the far end of the gradient with less risk control and increased red-teaming. Accessible interfaces with low-barrier sharing can also better enable cross-field collaborations \cite{DBLP:journals/corr/abs-1906-02569}. Large-scale probing can reveal flaws, as seen with Meta’s Galactica language model which was released with a demo. The demo was retracted within three days due to the public naming risks such as disinformation generation \cite{heaven_2022}. Both computer science training and low- to no-code interfaces are necessary to streamline sociotechnical research. Moral experiments show varying approaches to ethical problems by background and culture \cite{1334156}, which are urgently needed perspectives in building, evaluating, and deploying new AI systems. Effective design and user interface must be optimized for experts outside of computer science \cite{https://doi.org/10.48550/arxiv.1907.04446}.

\subsection{Closing Resource Gaps}
Resource gaps mainly between among major labs, research groups, and academia have widened \cite{benaich_hogarth_2022}. In addition to the gap hindering groups from developing systems at the same level of performance, it also hinders the ability to build and run exploratory research projects. The monetary, infrastructure, and sometimes skills limitations can bar especially underrepresented groups from contributing to understanding and mitigating risks. Public sector investment at national \cite{nairrtf_2021} and global levels can start to bridge this gap. Grants from developer labs can also sponsor third-party research, but should have built-in mechanisms for also allowing critical research. Infrastructure grants for computer clusters can enable smaller research groups to engage with powerful systems. Skill-building requires longer-term investment. 

\subsection{Technical and Practical Ethics Training}
Increasing access to social scientists and the many multidisciplinary experts underrepresented in the AI research community is insufficient. The technical barriers to evaluating and improving or mitigating harms of AI systems can slow or hinder critical research. Conversely, the lack of practical ethics and science and technology studies (STS) training among technical professionals prevents thoughtfully integrating societal guardrails from project conception throughout the development process. Training must be implemented at early stage education; academic courses and curricula in computer science must integrate social and ethical considerations. Social sciences geared toward examining AI systems must foster technical understanding. 

\subsection{Expert Foresight}
Experts in relevant disciplines should be included while relative risk is low. As generative AI systems become a higher risk for specific applications and fields such as disinformation and medical advice, correlating experts should be tapped in and begin foundational work in mitigating that risk. The rapid rate of AI development means substantial research should be planned in anticipation to prevent the trend of detection and mitigation trailing capabilities advancement. For example, research conducted on radicalization risks of language models using GPT-2 and GPT-3 show that GPT-3 has significantly higher risk potential in generating extremist text \cite{DBLP:journals/corr/abs-2009-06807}. Starting this research with less powerful systems can better inform future mitigation efforts.

\subsection{Multidisciplinary Discourse}
Increasing access to social scientists and the many multidisciplinary experts underrepresented in this discussion is insufficient. Critically, actors in this space must have some incentive to engage in frequent community discussion and be held accountable to commitments for safe releases. Google’s public position on responsible AI practices encourages in-house risk evaluation and mitigation, but conflicts of interest can result in internal critics being unable to share or publish findings \cite{10.1145/3514094.3534181} and dismissal \cite{ebell_2021}. These initiatives can also be formed as an industry argument for self-regulation, but ultimately lack external accountability. A new third-party convening body can help facilitate this discourse. Instead of relying on existing fora, conversations can take lessons from social and abolitionist movements in how to include underrepresented and affected communities \cite{stark_2021}. 

\subsection{Enforcement Mechanisms for Unsafe Release}
Ultimately all actors involved in the release of a powerful AI system must have some incentive to conduct releases safely. But enforcing responsible release requires a definition for what constitutes responsible release. Responsible is distinct from safe, and can emphasize meeting all possible enforceable safety guardrails pre- and post-release. Regulation can mandate that releases include system documentation and auditing for high-risk or high-impact releases. Updatable policies can recommend certain risk controls and guardrails and policy bodies can better fund risk research and development of further evaluations and controls. 

\section{Conclusion}
The gradient of generative AI system release shows the complexity and tradeoffs of any one option. Releases must balance concentration of power and AI risks in addition to considering precedent for future releases as system capabilities increase. Developers and deployers, regardless of release method preference, must engage multidisciplinary experts and the AI community to better form norms for safe release. Existing and evolving risk controls and guardrails require developer, deployer, researcher, and policymaker action and can mitigate some foreseeable harms, but long-term investments in disciplines and discourse across the AI community and among affected peoples are necessary.  

\begin{ack}
 Thank you to Hugging Face for funding this research. 

 Thank you to Joshua Achiam, Stella Biderman, Miles Brundage, Clémentine Fourrier, Yacine Jernite, Margaret Mitchell, Percy Liang, and Sonja Schmer-Galunder for their thoughtful feedback on earlier versions of this paper.

 Thank you to Johann Christensen for testing figures.

\end{ack}

\bibliography{release_gradient}

\begin{thebibliography}{108}
\providecommand{\natexlab}[1]{#1}
\providecommand{\url}[1]{\texttt{#1}}
\expandafter\ifx\csname urlstyle\endcsname\relax
  \providecommand{\doi}[1]{doi: #1}\else
  \providecommand{\doi}{doi: \begingroup \urlstyle{rm}\Url}\fi

\bibitem[Abid et~al.(2019)Abid, Abdalla, Abid, Khan, Alfozan, and
  Zou]{DBLP:journals/corr/abs-1906-02569}
A.~Abid, A.~Abdalla, A.~Abid, D.~Khan, A.~Alfozan, and J.~Y. Zou.
\newblock Gradio: Hassle-free sharing and testing of {ML} models in the wild.
\newblock \emph{CoRR}, abs/1906.02569, 2019.
\newblock URL \url{http://arxiv.org/abs/1906.02569}.

\bibitem[Alam et~al.(2022)Alam, Saha, Mukhopadhyay, and Kundu]{10.1145/3505634}
M.~Alam, S.~Saha, D.~Mukhopadhyay, and S.~Kundu.
\newblock Nn-lock: A lightweight authorization to prevent ip threats of deep
  learning models.
\newblock \emph{J. Emerg. Technol. Comput. Syst.}, 18\penalty0 (3), apr 2022.
\newblock ISSN 1550-4832.
\newblock \doi{10.1145/3505634}.
\newblock URL \url{https://doi.org/10.1145/3505634}.

\bibitem[Awad et~al.(2018)Awad, Dsouza, Kim, Schulz, Henrich, Shariff,
  Bonnefon, and Rahwan]{1334156}
E.~Awad, S.~Dsouza, R.~Kim, J.~Schulz, J.~Henrich, A.~Shariff, J.~F. Bonnefon,
  and I.~Rahwan.
\newblock The moral machine experiment.
\newblock \emph{Nature}, 563\penalty0 (7729):\penalty0 59--64, 2018.
\newblock URL \url{https://www.nature.com/articles/s41586-018-0637-6}.

\bibitem[Benaich and Hogarth(2022)]{benaich_hogarth_2022}
N.~Benaich and I.~Hogarth, 2022.
\newblock URL \url{https://www.stateof.ai/2022-report-launch.html}.

\bibitem[Bender and Friedman(2018)]{bender-friedman-2018-data}
E.~M. Bender and B.~Friedman.
\newblock Data statements for natural language processing: Toward mitigating
  system bias and enabling better science.
\newblock \emph{Transactions of the Association for Computational Linguistics},
  6:\penalty0 587--604, 2018.
\newblock \doi{10.1162/tacl_a_00041}.
\newblock URL \url{https://aclanthology.org/Q18-1041}.

\bibitem[Bender et~al.(2021)Bender, Gebru, McMillan-Major, and
  Shmitchell]{10.1145/3442188.3445922}
E.~M. Bender, T.~Gebru, A.~McMillan-Major, and S.~Shmitchell.
\newblock On the dangers of stochastic parrots: Can language models be too big?
  \includegraphics[height=1.2\fontcharht\font`\b]{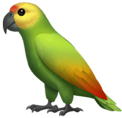}.
\newblock In \emph{Proceedings of the 2021 ACM Conference on Fairness,
  Accountability, and Transparency}, FAccT '21, page 610–623, New York, NY,
  USA, 2021. Association for Computing Machinery.
\newblock ISBN 9781450383097.
\newblock \doi{10.1145/3442188.3445922}.
\newblock URL \url{https://doi.org/10.1145/3442188.3445922}.

\bibitem[Benjamin(2020)]{benjamin_2020}
R.~Benjamin.
\newblock \emph{Race after technology: Abolitionist Tools for the new jim
  code}.
\newblock Polity, 2020.

\bibitem[Benkler(2019)]{benkler_2019}
Y.~Benkler.
\newblock Don't let industry write the rules for ai.
\newblock \emph{Nature News}, May 2019.
\newblock URL \url{https://www.nature.com/articles/d41586-019-01413-1}.

\bibitem[Bergman et~al.(2022)Bergman, Abercrombie, Spruit, Hovy, Dinan,
  Boureau, and Rieser]{bergman-etal-2022-guiding}
A.~S. Bergman, G.~Abercrombie, S.~Spruit, D.~Hovy, E.~Dinan, Y.-L. Boureau, and
  V.~Rieser.
\newblock Guiding the release of safer {E}2{E} conversational {AI} through
  value sensitive design.
\newblock In \emph{Proceedings of the 23rd Annual Meeting of the Special
  Interest Group on Discourse and Dialogue}, pages 39--52, Edinburgh, UK, Sept.
  2022. Association for Computational Linguistics.
\newblock URL \url{https://aclanthology.org/2022.sigdial-1.4}.

\bibitem[BigScience(2022)]{bigscience_2022}
BigScience.
\newblock Bigscience rail license v1.0, 2022.
\newblock URL \url{https://huggingface.co/spaces/bigscience/license}.

\bibitem[Birhane et~al.(2022)Birhane, Kalluri, Card, Agnew, Dotan, and
  Bao]{10.1145/3531146.3533083}
A.~Birhane, P.~Kalluri, D.~Card, W.~Agnew, R.~Dotan, and M.~Bao.
\newblock The values encoded in machine learning research.
\newblock In \emph{2022 ACM Conference on Fairness, Accountability, and
  Transparency}, FAccT '22, page 173–184, New York, NY, USA, 2022.
  Association for Computing Machinery.
\newblock ISBN 9781450393522.
\newblock \doi{10.1145/3531146.3533083}.
\newblock URL \url{https://doi.org/10.1145/3531146.3533083}.

\bibitem[Black et~al.(2022)Black, Biderman, Hallahan, Anthony, Gao, Golding,
  He, Leahy, McDonell, Phang, Pieler, Prashanth, Purohit, Reynolds, Tow, Wang,
  and Weinbach]{gpt-neox-20b}
S.~Black, S.~Biderman, E.~Hallahan, Q.~Anthony, L.~Gao, L.~Golding, H.~He,
  C.~Leahy, K.~McDonell, J.~Phang, M.~Pieler, U.~S. Prashanth, S.~Purohit,
  L.~Reynolds, J.~Tow, B.~Wang, and S.~Weinbach.
\newblock {GPT-NeoX-20B}: An open-source autoregressive language model.
\newblock In \emph{Proceedings of the ACL Workshop on Challenges \&
  Perspectives in Creating Large Language Models}, 2022.
\newblock URL \url{https://arxiv.org/abs/2204.06745}.

\bibitem[Blodgett et~al.(2020)Blodgett, Barocas, Daum{\'e}~III, and
  Wallach]{blodgett-etal-2020-language}
S.~L. Blodgett, S.~Barocas, H.~Daum{\'e}~III, and H.~Wallach.
\newblock Language (technology) is power: A critical survey of {``}bias{''} in
  {NLP}.
\newblock In \emph{Proceedings of the 58th Annual Meeting of the Association
  for Computational Linguistics}, pages 5454--5476, Online, July 2020.
  Association for Computational Linguistics.
\newblock \doi{10.18653/v1/2020.acl-main.485}.
\newblock URL \url{https://aclanthology.org/2020.acl-main.485}.

\bibitem[Boenisch(2021)]{10.3389/fdata.2021.729663}
F.~Boenisch.
\newblock A systematic review on model watermarking for neural networks.
\newblock \emph{Frontiers in Big Data}, 4, 2021.
\newblock ISSN 2624-909X.
\newblock \doi{10.3389/fdata.2021.729663}.
\newblock URL
  \url{https://www.frontiersin.org/articles/10.3389/fdata.2021.729663}.

\bibitem[Bommasani et~al.(2021)Bommasani, Hudson, Adeli, Altman, Arora, von
  Arx, Bernstein, Bohg, Bosselut, Brunskill, Brynjolfsson, Buch, Card,
  Castellon, Chatterji, Chen, Creel, Davis, Demszky, Donahue, Doumbouya,
  Durmus, Ermon, Etchemendy, Ethayarajh, Fei{-}Fei, Finn, Gale, Gillespie,
  Goel, Goodman, Grossman, Guha, Hashimoto, Henderson, Hewitt, Ho, Hong, Hsu,
  Huang, Icard, Jain, Jurafsky, Kalluri, Karamcheti, Keeling, Khani, Khattab,
  Koh, Krass, Krishna, Kuditipudi, and
  et~al.]{DBLP:journals/corr/abs-2108-07258}
R.~Bommasani, D.~A. Hudson, E.~Adeli, R.~Altman, S.~Arora, S.~von Arx, M.~S.
  Bernstein, J.~Bohg, A.~Bosselut, E.~Brunskill, E.~Brynjolfsson, S.~Buch,
  D.~Card, R.~Castellon, N.~S. Chatterji, A.~S. Chen, K.~Creel, J.~Q. Davis,
  D.~Demszky, C.~Donahue, M.~Doumbouya, E.~Durmus, S.~Ermon, J.~Etchemendy,
  K.~Ethayarajh, L.~Fei{-}Fei, C.~Finn, T.~Gale, L.~Gillespie, K.~Goel, N.~D.
  Goodman, S.~Grossman, N.~Guha, T.~Hashimoto, P.~Henderson, J.~Hewitt, D.~E.
  Ho, J.~Hong, K.~Hsu, J.~Huang, T.~Icard, S.~Jain, D.~Jurafsky, P.~Kalluri,
  S.~Karamcheti, G.~Keeling, F.~Khani, O.~Khattab, P.~W. Koh, M.~S. Krass,
  R.~Krishna, R.~Kuditipudi, and et~al.
\newblock On the opportunities and risks of foundation models.
\newblock \emph{CoRR}, abs/2108.07258, 2021.
\newblock URL \url{https://arxiv.org/abs/2108.07258}.

\bibitem[Brockman et~al.(2020)Brockman, Murati, Welinder, and
  OpenAI]{openai_2020}
G.~Brockman, M.~Murati, P.~Welinder, and OpenAI.
\newblock Openai api, Jun 2020.
\newblock URL \url{https://openai.com/blog/openai-api/}.

\bibitem[Brundage et~al.(2018)Brundage, Avin, Clark, Toner, Eckersley,
  Garfinkel, Dafoe, Scharre, Zeitzoff, Filar, Anderson, Roff, Allen,
  Steinhardt, Flynn, h{\'{E}}igeartaigh, Beard, Belfield, Farquhar, Lyle,
  Crootof, Evans, Page, Bryson, Yampolskiy, and
  Amodei]{DBLP:journals/corr/abs-1802-07228}
M.~Brundage, S.~Avin, J.~Clark, H.~Toner, P.~Eckersley, B.~Garfinkel, A.~Dafoe,
  P.~Scharre, T.~Zeitzoff, B.~Filar, H.~S. Anderson, H.~Roff, G.~C. Allen,
  J.~Steinhardt, C.~Flynn, S.~{\'{O}}. h{\'{E}}igeartaigh, S.~Beard,
  H.~Belfield, S.~Farquhar, C.~Lyle, R.~Crootof, O.~Evans, M.~Page, J.~Bryson,
  R.~Yampolskiy, and D.~Amodei.
\newblock The malicious use of artificial intelligence: Forecasting,
  prevention, and mitigation.
\newblock \emph{CoRR}, abs/1802.07228, 2018.
\newblock URL \url{http://arxiv.org/abs/1802.07228}.

\bibitem[Brundage et~al.(2020)Brundage, Avin, Wang, Belfield, Krueger,
  Hadfield, Khlaaf, Yang, Toner, Fong, Maharaj, Koh, Hooker, Leung, Trask,
  Bluemke, Lebensold, O'Keefe, Koren, Ryffel, Rubinovitz, Besiroglu, Carugati,
  Clark, Eckersley, de~Haas, Johnson, Laurie, Ingerman, Krawczuk, Askell,
  Cammarota, Lohn, Krueger, Stix, Henderson, Graham, Prunkl, Martin, Seger,
  Zilberman, hÉigeartaigh, Kroeger, Sastry, Kagan, Weller, Tse, Barnes, Dafoe,
  Scharre, Herbert-Voss, Rasser, Sodhani, Flynn, Gilbert, Dyer, Khan, Bengio,
  and Anderljung]{https://doi.org/10.48550/arxiv.2004.07213}
M.~Brundage, S.~Avin, J.~Wang, H.~Belfield, G.~Krueger, G.~Hadfield, H.~Khlaaf,
  J.~Yang, H.~Toner, R.~Fong, T.~Maharaj, P.~W. Koh, S.~Hooker, J.~Leung,
  A.~Trask, E.~Bluemke, J.~Lebensold, C.~O'Keefe, M.~Koren, T.~Ryffel,
  J.~Rubinovitz, T.~Besiroglu, F.~Carugati, J.~Clark, P.~Eckersley, S.~de~Haas,
  M.~Johnson, B.~Laurie, A.~Ingerman, I.~Krawczuk, A.~Askell, R.~Cammarota,
  A.~Lohn, D.~Krueger, C.~Stix, P.~Henderson, L.~Graham, C.~Prunkl, B.~Martin,
  E.~Seger, N.~Zilberman, S.~{\'O}. hÉigeartaigh, F.~Kroeger, G.~Sastry,
  R.~Kagan, A.~Weller, B.~Tse, E.~Barnes, A.~Dafoe, P.~Scharre,
  A.~Herbert-Voss, M.~Rasser, S.~Sodhani, C.~Flynn, T.~K. Gilbert, L.~Dyer,
  S.~Khan, Y.~Bengio, and M.~Anderljung.
\newblock Toward trustworthy ai development: Mechanisms for supporting
  verifiable claims, 2020.
\newblock URL \url{https://arxiv.org/abs/2004.07213}.

\bibitem[Cai and Zhu(2015)]{Cai2015TheCO}
L.~Cai and Y.~Zhu.
\newblock The challenges of data quality and data quality assessment in the big
  data era.
\newblock \emph{Data Sci. J.}, 14:\penalty0 2, 2015.

\bibitem[Caliskan et~al.(2017)Caliskan, Bryson, and
  Narayanan]{doi:10.1126/science.aal4230}
A.~Caliskan, J.~J. Bryson, and A.~Narayanan.
\newblock Semantics derived automatically from language corpora contain
  human-like biases.
\newblock \emph{Science}, 356\penalty0 (6334):\penalty0 183--186, 2017.
\newblock \doi{10.1126/science.aal4230}.
\newblock URL \url{https://www.science.org/doi/abs/10.1126/science.aal4230}.

\bibitem[Carlini et~al.(2020)Carlini, Tram{\`{e}}r, Wallace, Jagielski,
  Herbert{-}Voss, Lee, Roberts, Brown, Song, Erlingsson, Oprea, and
  Raffel]{DBLP:journals/corr/abs-2012-07805}
N.~Carlini, F.~Tram{\`{e}}r, E.~Wallace, M.~Jagielski, A.~Herbert{-}Voss,
  K.~Lee, A.~Roberts, T.~B. Brown, D.~Song, {\'{U}}.~Erlingsson, A.~Oprea, and
  C.~Raffel.
\newblock Extracting training data from large language models.
\newblock \emph{CoRR}, abs/2012.07805, 2020.
\newblock URL \url{https://arxiv.org/abs/2012.07805}.

\bibitem[CarperAI(2022)]{carperai_2022}
CarperAI.
\newblock Carperai, an eleutherai lab, announces plans for the first
  open-source "instruction-tuned" language model., Oct 2022.
\newblock URL \url{https://carper.ai/instruct-gpt-announcement/}.

\bibitem[Chakraborty et~al.(2020)Chakraborty, Mondal, and
  Srivastava]{10.5555/3437539.3437711}
A.~Chakraborty, A.~Mondal, and A.~Srivastava.
\newblock Hardware-assisted intellectual property protection of deep learning
  models.
\newblock In \emph{Proceedings of the 57th ACM/EDAC/IEEE Design Automation
  Conference}, DAC '20. IEEE Press, 2020.
\newblock ISBN 9781450367257.

\bibitem[Chowdhery et~al.(2022)Chowdhery, Narang, Devlin, Bosma, Mishra,
  Roberts, Barham, Chung, Sutton, Gehrmann, Schuh, Shi, Tsvyashchenko, Maynez,
  Rao, Barnes, Tay, Shazeer, Prabhakaran, Reif, Du, Hutchinson, Pope, Bradbury,
  Austin, Isard, Gur-Ari, Yin, Duke, Levskaya, Ghemawat, Dev, Michalewski,
  Garcia, Misra, Robinson, Fedus, Zhou, Ippolito, Luan, Lim, Zoph, Spiridonov,
  Sepassi, Dohan, Agrawal, Omernick, Dai, Pillai, Pellat, Lewkowycz, Moreira,
  Child, Polozov, Lee, Zhou, Wang, Saeta, Diaz, Firat, Catasta, Wei,
  Meier-Hellstern, Eck, Dean, Petrov, and
  Fiedel]{https://doi.org/10.48550/arxiv.2204.02311}
A.~Chowdhery, S.~Narang, J.~Devlin, M.~Bosma, G.~Mishra, A.~Roberts, P.~Barham,
  H.~W. Chung, C.~Sutton, S.~Gehrmann, P.~Schuh, K.~Shi, S.~Tsvyashchenko,
  J.~Maynez, A.~Rao, P.~Barnes, Y.~Tay, N.~Shazeer, V.~Prabhakaran, E.~Reif,
  N.~Du, B.~Hutchinson, R.~Pope, J.~Bradbury, J.~Austin, M.~Isard, G.~Gur-Ari,
  P.~Yin, T.~Duke, A.~Levskaya, S.~Ghemawat, S.~Dev, H.~Michalewski, X.~Garcia,
  V.~Misra, K.~Robinson, L.~Fedus, D.~Zhou, D.~Ippolito, D.~Luan, H.~Lim,
  B.~Zoph, A.~Spiridonov, R.~Sepassi, D.~Dohan, S.~Agrawal, M.~Omernick, A.~M.
  Dai, T.~S. Pillai, M.~Pellat, A.~Lewkowycz, E.~Moreira, R.~Child, O.~Polozov,
  K.~Lee, Z.~Zhou, X.~Wang, B.~Saeta, M.~Diaz, O.~Firat, M.~Catasta, J.~Wei,
  K.~Meier-Hellstern, D.~Eck, J.~Dean, S.~Petrov, and N.~Fiedel.
\newblock Palm: Scaling language modeling with pathways, 2022.
\newblock URL \url{https://arxiv.org/abs/2204.02311}.

\bibitem[Contractor et~al.(2020)Contractor, McDuff, Haines, Lee, Hines, and
  Hecht]{DBLP:journals/corr/abs-2011-03116}
D.~Contractor, D.~McDuff, J.~K. Haines, J.~Lee, C.~Hines, and B.~J. Hecht.
\newblock Behavioral use licensing for responsible {AI}.
\newblock \emph{CoRR}, abs/2011.03116, 2020.
\newblock URL \url{https://arxiv.org/abs/2011.03116}.

\bibitem[Crisan et~al.(2022)Crisan, Drouhard, Vig, and Rajani]{Crisan_2022}
A.~Crisan, M.~Drouhard, J.~Vig, and N.~Rajani.
\newblock Interactive model cards: A human-centered approach to model
  documentation.
\newblock In \emph{2022 {ACM} Conference on Fairness, Accountability, and
  Transparency}. {ACM}, jun 2022.
\newblock \doi{10.1145/3531146.3533108}.
\newblock URL \url{https://doi.org/10.1145%2F3531146.3533108}.

\bibitem[Dabbish et~al.(2012)Dabbish, Stuart, Tsay, and
  Herbsleb]{10.1145/2145204.2145396}
L.~Dabbish, C.~Stuart, J.~Tsay, and J.~Herbsleb.
\newblock Social coding in github: Transparency and collaboration in an open
  software repository.
\newblock In \emph{Proceedings of the ACM 2012 Conference on Computer Supported
  Cooperative Work}, CSCW '12, page 1277–1286, New York, NY, USA, 2012.
  Association for Computing Machinery.
\newblock ISBN 9781450310864.
\newblock \doi{10.1145/2145204.2145396}.
\newblock URL \url{https://doi.org/10.1145/2145204.2145396}.

\bibitem[Dave and Dastin(2020)]{dave_dastin_2020}
P.~Dave and J.~Dastin.
\newblock Google told its scientists to 'strike a positive tone' in ai research
  - documents.
\newblock \emph{Reuters}, 2020.
\newblock URL
  \url{https://www.reuters.com/article/us-alphabet-google-research-focus/google-told-its-scientists-to-strike-a-positive-tone-in-ai-research-documents-idUSKBN28X1CB}.

\bibitem[Dou et~al.(2021)Dou, Forbes, Koncel{-}Kedziorski, Smith, and
  Choi]{DBLP:journals/corr/abs-2107-01294}
Y.~Dou, M.~Forbes, R.~Koncel{-}Kedziorski, N.~A. Smith, and Y.~Choi.
\newblock Scarecrow: {A} framework for scrutinizing machine text.
\newblock \emph{CoRR}, abs/2107.01294, 2021.
\newblock URL \url{https://arxiv.org/abs/2107.01294}.

\bibitem[Ebell et~al.(2021)Ebell, Baeza-Yates, Benjamins, Cai, Coeckelbergh,
  Duarte, Hickok, Jacquet, Kim, Krijger, Macintyre, Madhamshettiwar, Maffeo,
  Matthews, Medsker, Smith, and Thais]{ebell_2021}
C.~Ebell, R.~Baeza-Yates, R.~Benjamins, H.~Cai, M.~Coeckelbergh, T.~Duarte,
  M.~Hickok, A.~Jacquet, A.~Kim, J.~Krijger, J.~Macintyre, P.~Madhamshettiwar,
  L.~Maffeo, J.~Matthews, L.~Medsker, P.~Smith, and S.~Thais.
\newblock Towards intellectual freedom in an ai ethics global community.
\newblock \emph{AI and Ethics}, 1, 04 2021.
\newblock \doi{10.1007/s43681-021-00052-5}.

\bibitem[Fan et~al.(2022)Fan, Ilic, Wolf, and
  Gall{\'e}]{bigscience-2022-bigscience}
A.~Fan, S.~Ilic, T.~Wolf, and M.~Gall{\'e}, editors.
\newblock \emph{Proceedings of BigScience Episode {\#}5 -- Workshop on
  Challenges {\&} Perspectives in Creating Large Language Models},
  virtual+Dublin, May 2022. Association for Computational Linguistics.
\newblock URL \url{https://aclanthology.org/2022.bigscience-1.0}.

\bibitem[Gao(2021)]{gao_2021}
L.~Gao.
\newblock On the sizes of openai api models, May 2021.
\newblock URL \url{https://blog.eleuther.ai/gpt3-model-sizes/}.

\bibitem[Gao et~al.(2021)Gao, Biderman, Black, Golding, Hoppe, Foster, Phang,
  He, Thite, Nabeshima, Presser, and Leahy]{DBLP:journals/corr/abs-2101-00027}
L.~Gao, S.~Biderman, S.~Black, L.~Golding, T.~Hoppe, C.~Foster, J.~Phang,
  H.~He, A.~Thite, N.~Nabeshima, S.~Presser, and C.~Leahy.
\newblock The pile: An 800gb dataset of diverse text for language modeling.
\newblock \emph{CoRR}, abs/2101.00027, 2021.
\newblock URL \url{https://arxiv.org/abs/2101.00027}.

\bibitem[Gebru et~al.(2018)Gebru, Morgenstern, Vecchione, Vaughan, Wallach,
  III, and Crawford]{DBLP:journals/corr/abs-1803-09010}
T.~Gebru, J.~Morgenstern, B.~Vecchione, J.~W. Vaughan, H.~M. Wallach, H.~D.
  III, and K.~Crawford.
\newblock Datasheets for datasets.
\newblock \emph{CoRR}, abs/1803.09010, 2018.
\newblock URL \url{http://arxiv.org/abs/1803.09010}.

\bibitem[Gehrmann et~al.(2019)Gehrmann, Strobelt, and
  Rush]{gehrmann-etal-2019-gltr}
S.~Gehrmann, H.~Strobelt, and A.~Rush.
\newblock {GLTR}: Statistical detection and visualization of generated text.
\newblock In \emph{Proceedings of the 57th Annual Meeting of the Association
  for Computational Linguistics: System Demonstrations}, pages 111--116,
  Florence, Italy, July 2019. Association for Computational Linguistics.
\newblock \doi{10.18653/v1/P19-3019}.

\bibitem[Gillespie(2021)]{gillespie_2021}
T.~Gillespie.
\newblock \emph{Custodians of the internet: Platforms, content moderation, and
  the hidden decisions that shape social media}.
\newblock Yale University Press, 2021.

\bibitem[Goldstein et~al.(2023)Goldstein, Sastry, Musser, DiResta, Gentzel, and
  Sedova]{https://doi.org/10.48550/arxiv.2301.04246}
J.~A. Goldstein, G.~Sastry, M.~Musser, R.~DiResta, M.~Gentzel, and K.~Sedova.
\newblock Generative language models and automated influence operations:
  Emerging threats and potential mitigations, 2023.
\newblock URL \url{https://arxiv.org/abs/2301.04246}.

\bibitem[Gragnaniello et~al.(2022)Gragnaniello, Marra, and
  Verdoliva]{Gragnaniello2022}
D.~Gragnaniello, F.~Marra, and L.~Verdoliva.
\newblock \emph{Detection of AI-Generated Synthetic Faces}, pages 191--212.
\newblock Springer International Publishing, Cham, 2022.
\newblock ISBN 978-3-030-87664-7.
\newblock \doi{10.1007/978-3-030-87664-7_9}.
\newblock URL \url{https://doi.org/10.1007/978-3-030-87664-7_9}.

\bibitem[Haimson et~al.(2021)Haimson, Delmonaco, Nie, and
  Wegner]{10.1145/3479610}
O.~L. Haimson, D.~Delmonaco, P.~Nie, and A.~Wegner.
\newblock Disproportionate removals and differing content moderation
  experiences for conservative, transgender, and black social media users:
  Marginalization and moderation gray areas.
\newblock \emph{Proc. ACM Hum.-Comput. Interact.}, 5\penalty0 (CSCW2), oct
  2021.
\newblock \doi{10.1145/3479610}.
\newblock URL \url{https://doi.org/10.1145/3479610}.

\bibitem[Heaven(2022)]{heaven_2022}
W.~D. Heaven.
\newblock Why meta's latest large language model survived only three days
  online, Nov 2022.
\newblock URL
  \url{https://www.technologyreview.com/2022/11/18/1063487/meta-large-language-model-ai-only-survived-three-days-gpt-3-science/}.

\bibitem[Hovy and Spruit(2016)]{hovy-spruit-2016-social}
D.~Hovy and S.~L. Spruit.
\newblock The social impact of natural language processing.
\newblock In \emph{Proceedings of the 54th Annual Meeting of the Association
  for Computational Linguistics (Volume 2: Short Papers)}, pages 591--598,
  Berlin, Germany, Aug. 2016. Association for Computational Linguistics.
\newblock \doi{10.18653/v1/P16-2096}.
\newblock URL \url{https://aclanthology.org/P16-2096}.

\bibitem[Jawahar et~al.(2020)Jawahar, Abdul{-}Mageed, and
  Lakshmanan]{DBLP:journals/corr/abs-2011-01314}
G.~Jawahar, M.~Abdul{-}Mageed, and L.~V.~S. Lakshmanan.
\newblock Automatic detection of machine generated text: {A} critical survey.
\newblock \emph{CoRR}, abs/2011.01314, 2020.
\newblock URL \url{https://arxiv.org/abs/2011.01314}.

\bibitem[Kalluri(2020)]{Kalluri_2020}
P.~Kalluri.
\newblock Don’t ask if artificial intelligence is good or fair, ask how it
  shifts power.
\newblock \emph{Nature News}, July 2020.
\newblock URL \url{https://www.nature.com/articles/d41586-020-02003-2}.

\bibitem[Kaloudi and Li(2020)]{10.1145/3372823}
N.~Kaloudi and J.~Li.
\newblock The ai-based cyber threat landscape: A survey.
\newblock \emph{ACM Comput. Surv.}, 53\penalty0 (1), feb 2020.
\newblock ISSN 0360-0300.
\newblock \doi{10.1145/3372823}.
\newblock URL \url{https://doi.org/10.1145/3372823}.

\bibitem[Khlaaf et~al.(2022)Khlaaf, Mishkin, Achiam, Krueger, and
  Brundage]{https://doi.org/10.48550/arxiv.2207.14157}
H.~Khlaaf, P.~Mishkin, J.~Achiam, G.~Krueger, and M.~Brundage.
\newblock A hazard analysis framework for code synthesis large language models,
  2022.
\newblock URL \url{https://arxiv.org/abs/2207.14157}.

\bibitem[Kirchenbauer et~al.(2023)Kirchenbauer, Geiping, Wen, Katz, Miers, and
  Goldstein]{https://doi.org/10.48550/arxiv.2301.10226}
J.~Kirchenbauer, J.~Geiping, Y.~Wen, J.~Katz, I.~Miers, and T.~Goldstein.
\newblock A watermark for large language models, 2023.
\newblock URL \url{https://arxiv.org/abs/2301.10226}.

\bibitem[Kreps et~al.(2022)Kreps, McCain, and
  Brundage]{kreps_mccain_brundage_2022}
S.~Kreps, R.~M. McCain, and M.~Brundage.
\newblock All the news that’s fit to fabricate: Ai-generated text as a tool
  of media misinformation.
\newblock \emph{Journal of Experimental Political Science}, 9\penalty0
  (1):\penalty0 104–117, 2022.
\newblock \doi{10.1017/XPS.2020.37}.

\bibitem[Liang et~al.(2022{\natexlab{a}})Liang, Bommasani, Creel, and
  Reich]{liang_bommasani_creel_reich}
P.~Liang, R.~Bommasani, K.~Creel, and R.~Reich.
\newblock The time is now to develop community norms for the release of
  foundation models, 2022{\natexlab{a}}.
\newblock URL
  \url{https://hai.stanford.edu/news/time-now-develop-community-norms-release-foundation-models}.

\bibitem[Liang et~al.(2022{\natexlab{b}})Liang, Bommasani, Lee, Tsipras, Soylu,
  Yasunaga, Zhang, Narayanan, Wu, Kumar, Newman, Yuan, Yan, Zhang, Cosgrove,
  Manning, Ré, Acosta-Navas, Hudson, Zelikman, Durmus, Ladhak, Rong, Ren, Yao,
  Wang, Santhanam, Orr, Zheng, Yuksekgonul, Suzgun, Kim, Guha, Chatterji,
  Khattab, Henderson, Huang, Chi, Xie, Santurkar, Ganguli, Hashimoto, Icard,
  Zhang, Chaudhary, Wang, Li, Mai, Zhang, and
  Koreeda]{https://doi.org/10.48550/arxiv.2211.09110}
P.~Liang, R.~Bommasani, T.~Lee, D.~Tsipras, D.~Soylu, M.~Yasunaga, Y.~Zhang,
  D.~Narayanan, Y.~Wu, A.~Kumar, B.~Newman, B.~Yuan, B.~Yan, C.~Zhang,
  C.~Cosgrove, C.~D. Manning, C.~Ré, D.~Acosta-Navas, D.~A. Hudson,
  E.~Zelikman, E.~Durmus, F.~Ladhak, F.~Rong, H.~Ren, H.~Yao, J.~Wang,
  K.~Santhanam, L.~Orr, L.~Zheng, M.~Yuksekgonul, M.~Suzgun, N.~Kim, N.~Guha,
  N.~Chatterji, O.~Khattab, P.~Henderson, Q.~Huang, R.~Chi, S.~M. Xie,
  S.~Santurkar, S.~Ganguli, T.~Hashimoto, T.~Icard, T.~Zhang, V.~Chaudhary,
  W.~Wang, X.~Li, Y.~Mai, Y.~Zhang, and Y.~Koreeda.
\newblock Holistic evaluation of language models, 2022{\natexlab{b}}.
\newblock URL \url{https://arxiv.org/abs/2211.09110}.

\bibitem[Liao(2022)]{LiaoModelTracker2022}
T.~Liao.
\newblock Foundation model tracker.
\newblock https://foundationmodeltracker.com, 2022.

\bibitem[Lipton(2019)]{lipton_2019}
Z.~C. Lipton.
\newblock Openai trains language model, mass hysteria ensues, Feb 2019.
\newblock URL
  \url{https://www.approximatelycorrect.com/2019/02/17/openai-trains-language-model-mass-hysteria-ensues/}.

\bibitem[Mandel et~al.(2019)Mandel, Best, Tanaka, Temple, Haili, Schlectinger,
  and Szeto]{https://doi.org/10.48550/arxiv.1907.04446}
T.~Mandel, J.~Best, R.~H. Tanaka, H.~Temple, C.~Haili, K.~Schlectinger, and
  R.~Szeto.
\newblock Let's keep it safe: Designing user interfaces that allow everyone to
  contribute to ai safety, 2019.
\newblock URL \url{https://arxiv.org/abs/1907.04446}.

\bibitem[McGuffie and Newhouse(2020)]{DBLP:journals/corr/abs-2009-06807}
K.~McGuffie and A.~Newhouse.
\newblock The radicalization risks of {GPT-3} and advanced neural language
  models.
\newblock \emph{CoRR}, abs/2009.06807, 2020.
\newblock URL \url{https://arxiv.org/abs/2009.06807}.

\bibitem[Michael et~al.(2022)Michael, Holtzman, Parrish, Mueller, Wang, Chen,
  Madaan, Nangia, Pang, Phang, and
  Bowman]{https://doi.org/10.48550/arxiv.2208.12852}
J.~Michael, A.~Holtzman, A.~Parrish, A.~Mueller, A.~Wang, A.~Chen, D.~Madaan,
  N.~Nangia, R.~Y. Pang, J.~Phang, and S.~R. Bowman.
\newblock What do nlp researchers believe? results of the nlp community
  metasurvey, 2022.
\newblock URL \url{https://arxiv.org/abs/2208.12852}.

\bibitem[Microsoft(2022)]{microsoft_2022}
Microsoft.
\newblock Microsoft turing academic program (ms-tap), Oct 2022.
\newblock URL
  \url{https://www.microsoft.com/en-us/research/collaboration/microsoft-turing-academic-program/}.

\bibitem[Midjourney(2022)]{midjourney_2022}
Midjourney.
\newblock Quick start guide, 2022.
\newblock URL \url{https://midjourney.gitbook.io/docs/}.

\bibitem[Mishkin et~al.(2022)Mishkin, Ahmad, Brundage, Krueger, and
  Sastry]{mishkin2022risks}
P.~Mishkin, L.~Ahmad, M.~Brundage, G.~Krueger, and G.~Sastry.
\newblock Dall·e 2 preview - risks and limitations.
\newblock 2022.
\newblock URL
  \url{[https://github.com/openai/dalle-2-preview/blob/main/system-card.md](https://github.com/openai/dalle-2-preview/blob/main/system-card.md)}.

\bibitem[Mitchell et~al.(2018)Mitchell, Wu, Zaldivar, Barnes, Vasserman,
  Hutchinson, Spitzer, Raji, and Gebru]{DBLP:journals/corr/abs-1810-03993}
M.~Mitchell, S.~Wu, A.~Zaldivar, P.~Barnes, L.~Vasserman, B.~Hutchinson,
  E.~Spitzer, I.~D. Raji, and T.~Gebru.
\newblock Model cards for model reporting.
\newblock \emph{CoRR}, abs/1810.03993, 2018.
\newblock URL \url{http://arxiv.org/abs/1810.03993}.

\bibitem[Mohamed et~al.(2020)Mohamed, Png, and
  Isaac]{DBLP:journals/corr/abs-2007-04068}
S.~Mohamed, M.~Png, and W.~Isaac.
\newblock Decolonial {AI:} decolonial theory as sociotechnical foresight in
  artificial intelligence.
\newblock \emph{CoRR}, abs/2007.04068, 2020.
\newblock URL \url{https://arxiv.org/abs/2007.04068}.

\bibitem[Mostaque(2022)]{mostaque_2022}
E.~Mostaque.
\newblock Stable diffusion public release, Sep 2022.
\newblock URL \url{https://stability.ai/blog/stable-diffusion-public-release}.

\bibitem[NAIRRTF(2021)]{nairrtf_2021}
NAIRRTF.
\newblock The national artificial intelligence research resource task force
  (nairrtf), 2021.
\newblock URL \url{https://www.ai.gov/nairrtf/}.

\bibitem[Nangia et~al.(2020)Nangia, Vania, Bhalerao, and
  Bowman]{DBLP:journals/corr/abs-2010-00133}
N.~Nangia, C.~Vania, R.~Bhalerao, and S.~R. Bowman.
\newblock Crows-pairs: {A} challenge dataset for measuring social biases in
  masked language models.
\newblock \emph{CoRR}, abs/2010.00133, 2020.
\newblock URL \url{https://arxiv.org/abs/2010.00133}.

\bibitem[Nguyen et~al.(2019)Nguyen, Nguyen, Nguyen, Nguyen, and
  Nahavandi]{DBLP:journals/corr/abs-1909-11573}
T.~T. Nguyen, C.~M. Nguyen, D.~T. Nguyen, D.~T. Nguyen, and S.~Nahavandi.
\newblock Deep learning for deepfakes creation and detection.
\newblock \emph{CoRR}, abs/1909.11573, 2019.
\newblock URL \url{http://arxiv.org/abs/1909.11573}.

\bibitem[Noble(2018)]{noble_2018}
S.~U. Noble.
\newblock \emph{Algorithms of oppression how search engines reinforce racism}.
\newblock New York University Press, 2018.

\bibitem[OpenAI(2019)]{openai_2019}
OpenAI.
\newblock Gpt-2 model card, Nov 2019.
\newblock URL \url{https://github.com/openai/gpt-2/blob/master/model_card.md}.

\bibitem[OpenAI(2020)]{openai_gpt3}
OpenAI.
\newblock Gpt-3 model card, Sep 2020.
\newblock URL \url{https://github.com/openai/gpt-3/blob/master/model-card.md}.

\bibitem[OpenAI(2022{\natexlab{a}})]{openai_2022}
OpenAI.
\newblock Dall·e api now available in public beta, Nov 2022{\natexlab{a}}.
\newblock URL
  \url{https://openai.com/blog/dall-e-api-now-available-in-public-beta/}.

\bibitem[OpenAI(2022{\natexlab{b}})]{openai_chat}
OpenAI.
\newblock Chatgpt: Optimizing language models for dialogue, Nov
  2022{\natexlab{b}}.
\newblock URL \url{https://openai.com/blog/chatgpt/}.

\bibitem[OpenAI(2022{\natexlab{c}})]{openai_rate}
OpenAI.
\newblock What's the rate limit for the dall·e api? how can i request an
  increase?, 2022{\natexlab{c}}.
\newblock URL
  \url{https://help.openai.com/en/articles/6696591-what-s-the-rate-limit-for-the-dall-e-api-how-can-i-request-an-increase}.

\bibitem[Ouyang et~al.(2022)Ouyang, Wu, Jiang, Almeida, Wainwright, Mishkin,
  Zhang, Agarwal, Slama, Ray, Schulman, Hilton, Kelton, Miller, Simens, Askell,
  Welinder, Christiano, Leike, and
  Lowe]{https://doi.org/10.48550/arxiv.2203.02155}
L.~Ouyang, J.~Wu, X.~Jiang, D.~Almeida, C.~L. Wainwright, P.~Mishkin, C.~Zhang,
  S.~Agarwal, K.~Slama, A.~Ray, J.~Schulman, J.~Hilton, F.~Kelton, L.~Miller,
  M.~Simens, A.~Askell, P.~Welinder, P.~Christiano, J.~Leike, and R.~Lowe.
\newblock Training language models to follow instructions with human feedback,
  2022.
\newblock URL \url{https://arxiv.org/abs/2203.02155}.

\bibitem[Ovadya and Whittlestone(2019)]{DBLP:journals/corr/abs-1907-11274}
A.~Ovadya and J.~Whittlestone.
\newblock Reducing malicious use of synthetic media research: Considerations
  and potential release practices for machine learning.
\newblock \emph{CoRR}, abs/1907.11274, 2019.
\newblock URL \url{http://arxiv.org/abs/1907.11274}.

\bibitem[Procope et~al.(2022)Procope, Cheema, Adkins, Alsallakh, Green,
  McReynolds, Pehl, Wang, and Zvyagina]{Procope_2022}
C.~Procope, A.~Cheema, D.~Adkins, B.~Alsallakh, N.~Green, E.~McReynolds,
  G.~Pehl, E.~Wang, and P.~Zvyagina.
\newblock System-level transparency of machine learning.
\newblock 2022.
\newblock URL
  \url{https://ai.facebook.com/research/publications/system-level-transparency-of-machine-learning/}.

\bibitem[Rae et~al.(2021)Rae, Borgeaud, Cai, Millican, Hoffmann, Song,
  Aslanides, Henderson, Ring, Young, Rutherford, Hennigan, Menick, Cassirer,
  Powell, van~den Driessche, Hendricks, Rauh, Huang, Glaese, Welbl, Dathathri,
  Huang, Uesato, Mellor, Higgins, Creswell, McAleese, Wu, Elsen, Jayakumar,
  Buchatskaya, Budden, Sutherland, Simonyan, Paganini, Sifre, Martens, Li,
  Kuncoro, Nematzadeh, Gribovskaya, Donato, Lazaridou, Mensch, Lespiau,
  Tsimpoukelli, Grigorev, Fritz, Sottiaux, Pajarskas, Pohlen, Gong, Toyama,
  de~Masson~d'Autume, Li, Terzi, Mikulik, Babuschkin, Clark, de~Las~Casas, Guy,
  Jones, Bradbury, Johnson, Hechtman, Weidinger, Gabriel, Isaac, Lockhart,
  Osindero, Rimell, Dyer, Vinyals, Ayoub, Stanway, Bennett, Hassabis,
  Kavukcuoglu, and Irving]{DBLP:journals/corr/abs-2112-11446}
J.~W. Rae, S.~Borgeaud, T.~Cai, K.~Millican, J.~Hoffmann, H.~F. Song,
  J.~Aslanides, S.~Henderson, R.~Ring, S.~Young, E.~Rutherford, T.~Hennigan,
  J.~Menick, A.~Cassirer, R.~Powell, G.~van~den Driessche, L.~A. Hendricks,
  M.~Rauh, P.~Huang, A.~Glaese, J.~Welbl, S.~Dathathri, S.~Huang, J.~Uesato,
  J.~Mellor, I.~Higgins, A.~Creswell, N.~McAleese, A.~Wu, E.~Elsen, S.~M.
  Jayakumar, E.~Buchatskaya, D.~Budden, E.~Sutherland, K.~Simonyan,
  M.~Paganini, L.~Sifre, L.~Martens, X.~L. Li, A.~Kuncoro, A.~Nematzadeh,
  E.~Gribovskaya, D.~Donato, A.~Lazaridou, A.~Mensch, J.~Lespiau,
  M.~Tsimpoukelli, N.~Grigorev, D.~Fritz, T.~Sottiaux, M.~Pajarskas, T.~Pohlen,
  Z.~Gong, D.~Toyama, C.~de~Masson~d'Autume, Y.~Li, T.~Terzi, V.~Mikulik,
  I.~Babuschkin, A.~Clark, D.~de~Las~Casas, A.~Guy, C.~Jones, J.~Bradbury,
  M.~Johnson, B.~A. Hechtman, L.~Weidinger, I.~Gabriel, W.~S. Isaac,
  E.~Lockhart, S.~Osindero, L.~Rimell, C.~Dyer, O.~Vinyals, K.~Ayoub,
  J.~Stanway, L.~Bennett, D.~Hassabis, K.~Kavukcuoglu, and G.~Irving.
\newblock Scaling language models: Methods, analysis {\&} insights from
  training gopher.
\newblock \emph{CoRR}, abs/2112.11446, 2021.
\newblock URL \url{https://arxiv.org/abs/2112.11446}.

\bibitem[Raji et~al.(2020)Raji, Smart, White, Mitchell, Gebru, Hutchinson,
  Smith-Loud, Theron, and Barnes]{10.1145/3351095.3372873}
I.~D. Raji, A.~Smart, R.~N. White, M.~Mitchell, T.~Gebru, B.~Hutchinson,
  J.~Smith-Loud, D.~Theron, and P.~Barnes.
\newblock Closing the ai accountability gap: Defining an end-to-end framework
  for internal algorithmic auditing.
\newblock In \emph{Proceedings of the 2020 Conference on Fairness,
  Accountability, and Transparency}, FAT* '20, page 33–44, New York, NY, USA,
  2020. Association for Computing Machinery.
\newblock ISBN 9781450369367.
\newblock \doi{10.1145/3351095.3372873}.
\newblock URL \url{https://doi.org/10.1145/3351095.3372873}.

\bibitem[Raji et~al.(2021)Raji, Bender, Paullada, Denton, and
  Hanna]{DBLP:journals/corr/abs-2111-15366}
I.~D. Raji, E.~M. Bender, A.~Paullada, E.~Denton, and A.~Hanna.
\newblock {AI} and the everything in the whole wide world benchmark.
\newblock \emph{CoRR}, abs/2111.15366, 2021.
\newblock URL \url{https://arxiv.org/abs/2111.15366}.

\bibitem[Raji et~al.(2022)Raji, Xu, Honigsberg, and
  Ho]{10.1145/3514094.3534181}
I.~D. Raji, P.~Xu, C.~Honigsberg, and D.~Ho.
\newblock Outsider oversight: Designing a third party audit ecosystem for ai
  governance.
\newblock In \emph{Proceedings of the 2022 AAAI/ACM Conference on AI, Ethics,
  and Society}, AIES '22, page 557–571, New York, NY, USA, 2022. Association
  for Computing Machinery.
\newblock ISBN 9781450392471.
\newblock \doi{10.1145/3514094.3534181}.
\newblock URL \url{https://doi.org/10.1145/3514094.3534181}.

\bibitem[Ramesh et~al.(2022)Ramesh, Dhariwal, Nichol, Chu, and
  Chen]{https://doi.org/10.48550/arxiv.2204.06125}
A.~Ramesh, P.~Dhariwal, A.~Nichol, C.~Chu, and M.~Chen.
\newblock Hierarchical text-conditional image generation with clip latents,
  2022.
\newblock URL \url{https://arxiv.org/abs/2204.06125}.

\bibitem[Rando et~al.(2022)Rando, Paleka, Lindner, Heim, and
  Tramèr]{https://doi.org/10.48550/arxiv.2210.04610}
J.~Rando, D.~Paleka, D.~Lindner, L.~Heim, and F.~Tramèr.
\newblock Red-teaming the stable diffusion safety filter, 2022.
\newblock URL \url{https://arxiv.org/abs/2210.04610}.

\bibitem[Reich(2022)]{reich_2022}
R.~Reich.
\newblock Now ai can write students' essays for them, will everyone become a
  cheat?, Nov 2022.
\newblock URL
  \url{https://www.theguardian.com/commentisfree/2022/nov/28/ai-students-essays-cheat-teachers-plagiarism-tech}.

\bibitem[Rombach and Esser(2022)]{rombach_esser_2022}
R.~Rombach and P.~Esser.
\newblock Stable-diffusion license, Aug 2022.
\newblock URL
  \url{https://github.com/CompVis/stable-diffusion/blob/main/LICENSE}.

\bibitem[Rombach et~al.(2022)Rombach, Blattmann, Lorenz, Esser, and
  Ommer]{Rombach_2022_CVPR}
R.~Rombach, A.~Blattmann, D.~Lorenz, P.~Esser, and B.~Ommer.
\newblock High-resolution image synthesis with latent diffusion models.
\newblock In \emph{Proceedings of the IEEE/CVF Conference on Computer Vision
  and Pattern Recognition (CVPR)}, pages 10684--10695, June 2022.

\bibitem[Rubinovitz(2018)]{rubinovitz_2018}
J.~Rubinovitz.
\newblock Bias bounty programs as a method of combatting bias in ai, 2018.
\newblock URL
  \url{https://rubinovitz.com/2018/08/01/bias-bounty-programs-as-a-method-of-combatting/}.

\bibitem[Saharia et~al.(2022)Saharia, Chan, Saxena, Li, Whang, Denton,
  Ghasemipour, Ayan, Mahdavi, Lopes, Salimans, Ho, Fleet, and
  Norouzi]{https://doi.org/10.48550/arxiv.2205.11487}
C.~Saharia, W.~Chan, S.~Saxena, L.~Li, J.~Whang, E.~Denton, S.~K.~S.
  Ghasemipour, B.~K. Ayan, S.~S. Mahdavi, R.~G. Lopes, T.~Salimans, J.~Ho,
  D.~J. Fleet, and M.~Norouzi.
\newblock Photorealistic text-to-image diffusion models with deep language
  understanding, 2022.
\newblock URL \url{https://arxiv.org/abs/2205.11487}.

\bibitem[Sastry(2021)]{sastry}
G.~Sastry.
\newblock Beyond “release” vs. “not release”, 2021.
\newblock URL
  \url{https://crfm.stanford.edu/commentary/2021/10/18/sastry.html}.

\bibitem[Shelby et~al.(2022)Shelby, Rismani, Henne, Moon, Rostamzadeh,
  Nicholas, Yilla, Gallegos, Smart, Garcia, and
  Virk]{https://doi.org/10.48550/arxiv.2210.05791}
R.~Shelby, S.~Rismani, K.~Henne, A.~Moon, N.~Rostamzadeh, P.~Nicholas,
  N.~Yilla, J.~Gallegos, A.~Smart, E.~Garcia, and G.~Virk.
\newblock Sociotechnical harms: Scoping a taxonomy for harm reduction, 2022.
\newblock URL \url{https://arxiv.org/abs/2210.05791}.

\bibitem[Shevlane(2022)]{https://doi.org/10.48550/arxiv.2201.05159}
T.~Shevlane.
\newblock Structured access: an emerging paradigm for safe ai deployment, 2022.
\newblock URL \url{https://arxiv.org/abs/2201.05159}.

\bibitem[Shevlane and Dafoe(2020)]{DBLP:journals/corr/abs-2001-00463}
T.~Shevlane and A.~Dafoe.
\newblock The offense-defense balance of scientific knowledge: Does publishing
  {AI} research reduce misuse?
\newblock \emph{CoRR}, abs/2001.00463, 2020.
\newblock URL \url{http://arxiv.org/abs/2001.00463}.

\bibitem[Simonite(2021)]{simonite_2021}
T.~Simonite.
\newblock It began as an ai-fueled dungeon game. it got much darker, May 2021.
\newblock URL
  \url{https://www.wired.com/story/ai-fueled-dungeon-game-got-much-darker/}.

\bibitem[Singer et~al.(2022)Singer, Polyak, Hayes, Yin, An, Zhang, Hu, Yang,
  Ashual, Gafni, Parikh, Gupta, and
  Taigman]{https://doi.org/10.48550/arxiv.2209.14792}
U.~Singer, A.~Polyak, T.~Hayes, X.~Yin, J.~An, S.~Zhang, Q.~Hu, H.~Yang,
  O.~Ashual, O.~Gafni, D.~Parikh, S.~Gupta, and Y.~Taigman.
\newblock Make-a-video: Text-to-video generation without text-video data, 2022.
\newblock URL \url{https://arxiv.org/abs/2209.14792}.

\bibitem[Solaiman and Dennison(2021)]{solaiman2021process}
I.~Solaiman and C.~Dennison.
\newblock Process for adapting language models to society (palms) with
  values-targeted datasets, 2021.

\bibitem[Solaiman et~al.(2019)Solaiman, Brundage, Clark, Askell,
  Herbert{-}Voss, Wu, Radford, and Wang]{DBLP:journals/corr/abs-1908-09203}
I.~Solaiman, M.~Brundage, J.~Clark, A.~Askell, A.~Herbert{-}Voss, J.~Wu,
  A.~Radford, and J.~Wang.
\newblock Release strategies and the social impacts of language models.
\newblock \emph{CoRR}, abs/1908.09203, 2019.
\newblock URL \url{http://arxiv.org/abs/1908.09203}.

\bibitem[Staff(2022)]{staff_2022}
P.~Staff.
\newblock Publication norms for responsible ai, Dec 2022.
\newblock URL
  \url{https://partnershiponai.org/workstream/publication-norms-for-responsible-ai/}.

\bibitem[Stark et~al.(2021)Stark, Greene, and Hoffmann]{stark_2021}
L.~Stark, D.~Greene, and A.~Hoffmann.
\newblock \emph{Critical Perspectives on Governance Mechanisms for AI/ML
  Systems}, pages 257--280.
\newblock 01 2021.
\newblock ISBN 978-3-030-56285-4.
\newblock \doi{10.1007/978-3-030-56286-1_9}.

\bibitem[Talat et~al.(2022)Talat, N{\'e}v{\'e}ol, Biderman, Clinciu, Dey,
  Longpre, Luccioni, Masoud, Mitchell, Radev, Sharma, Subramonian, Tae, Tan,
  Tunuguntla, and Van Der~Wal]{talat-etal-2022-reap}
Z.~Talat, A.~N{\'e}v{\'e}ol, S.~Biderman, M.~Clinciu, M.~Dey, S.~Longpre,
  S.~Luccioni, M.~Masoud, M.~Mitchell, D.~Radev, S.~Sharma, A.~Subramonian,
  J.~Tae, S.~Tan, D.~Tunuguntla, and O.~Van Der~Wal.
\newblock You reap what you sow: On the challenges of bias evaluation under
  multilingual settings.
\newblock In \emph{Proceedings of BigScience Episode {\#}5 -- Workshop on
  Challenges {\&} Perspectives in Creating Large Language Models}, pages
  26--41, virtual+Dublin, May 2022. Association for Computational Linguistics.
\newblock \doi{10.18653/v1/2022.bigscience-1.3}.
\newblock URL \url{https://aclanthology.org/2022.bigscience-1.3}.

\bibitem[Thoppilan et~al.(2022)Thoppilan, De~Freitas, Hall, Shazeer,
  Kulshreshtha, Cheng, Jin, Bos, Baker, Du, Li, Lee, Zheng, Ghafouri, Menegali,
  Huang, Krikun, Lepikhin, Qin, Chen, Xu, Chen, Roberts, Bosma, Zhao, Zhou,
  Chang, Krivokon, Rusch, Pickett, Srinivasan, Man, Meier-Hellstern, Morris,
  Doshi, Santos, Duke, Soraker, Zevenbergen, Prabhakaran, Diaz, Hutchinson,
  Olson, Molina, Hoffman-John, Lee, Aroyo, Rajakumar, Butryna, Lamm, Kuzmina,
  Fenton, Cohen, Bernstein, Kurzweil, Aguera-Arcas, Cui, Croak, Chi, and
  Le]{https://doi.org/10.48550/arxiv.2201.08239}
R.~Thoppilan, D.~De~Freitas, J.~Hall, N.~Shazeer, A.~Kulshreshtha, H.-T. Cheng,
  A.~Jin, T.~Bos, L.~Baker, Y.~Du, Y.~Li, H.~Lee, H.~S. Zheng, A.~Ghafouri,
  M.~Menegali, Y.~Huang, M.~Krikun, D.~Lepikhin, J.~Qin, D.~Chen, Y.~Xu,
  Z.~Chen, A.~Roberts, M.~Bosma, V.~Zhao, Y.~Zhou, C.-C. Chang, I.~Krivokon,
  W.~Rusch, M.~Pickett, P.~Srinivasan, L.~Man, K.~Meier-Hellstern, M.~R.
  Morris, T.~Doshi, R.~D. Santos, T.~Duke, J.~Soraker, B.~Zevenbergen,
  V.~Prabhakaran, M.~Diaz, B.~Hutchinson, K.~Olson, A.~Molina, E.~Hoffman-John,
  J.~Lee, L.~Aroyo, R.~Rajakumar, A.~Butryna, M.~Lamm, V.~Kuzmina, J.~Fenton,
  A.~Cohen, R.~Bernstein, R.~Kurzweil, B.~Aguera-Arcas, C.~Cui, M.~Croak,
  E.~Chi, and Q.~Le.
\newblock Lamda: Language models for dialog applications, 2022.
\newblock URL \url{https://arxiv.org/abs/2201.08239}.

\bibitem[Trajtenberg(2018)]{Trajtenberg18}
M.~Trajtenberg.
\newblock Ai as the next gpt: a political-economy perspective.
\newblock Working Paper 24245, National Bureau of Economic Research, January
  2018.
\newblock URL \url{http://www.nber.org/papers/w24245}.

\bibitem[Varshney et~al.(2020)Varshney, Keskar, and
  Socher]{DBLP:journals/corr/abs-2002-03438}
L.~R. Varshney, N.~S. Keskar, and R.~Socher.
\newblock Limits of detecting text generated by large-scale language models.
\newblock \emph{CoRR}, abs/2002.03438, 2020.
\newblock URL \url{https://arxiv.org/abs/2002.03438}.

\bibitem[Vee(2022)]{vee_2022}
A.~Vee.
\newblock Automated trolling: The case of gpt-4chan when artificial
  intelligence is as easy as writing.
\newblock \emph{Interfaces: Essays and Reviews in Computing and Culture Vol. 3,
  Charles Babbage Institute, University of Minnesota}, pages 102--111, 2022.

\bibitem[Wang(2021)]{mesh-transformer-jax}
B.~Wang.
\newblock {Mesh-Transformer-JAX: Model-Parallel Implementation of Transformer
  Language Model with JAX}.
\newblock \url{https://github.com/kingoflolz/mesh-transformer-jax}, May 2021.

\bibitem[Weidinger et~al.(2021)Weidinger, Mellor, Rauh, Griffin, Uesato, Huang,
  Cheng, Glaese, Balle, Kasirzadeh, Kenton, Brown, Hawkins, Stepleton, Biles,
  Birhane, Haas, Rimell, Hendricks, Isaac, Legassick, Irving, and
  Gabriel]{https://doi.org/10.48550/arxiv.2112.04359}
L.~Weidinger, J.~Mellor, M.~Rauh, C.~Griffin, J.~Uesato, P.-S. Huang, M.~Cheng,
  M.~Glaese, B.~Balle, A.~Kasirzadeh, Z.~Kenton, S.~Brown, W.~Hawkins,
  T.~Stepleton, C.~Biles, A.~Birhane, J.~Haas, L.~Rimell, L.~A. Hendricks,
  W.~Isaac, S.~Legassick, G.~Irving, and I.~Gabriel.
\newblock Ethical and social risks of harm from language models, 2021.
\newblock URL \url{https://arxiv.org/abs/2112.04359}.

\bibitem[Whittaker(2021)]{10.1145/3488666}
M.~Whittaker.
\newblock The steep cost of capture.
\newblock \emph{Interactions}, 28\penalty0 (6):\penalty0 50–55, nov 2021.
\newblock ISSN 1072-5520.
\newblock \doi{10.1145/3488666}.
\newblock URL \url{https://doi.org/10.1145/3488666}.

\bibitem[Whittlestone and Ovadya(2019)]{DBLP:journals/corr/abs-1910-01170}
J.~Whittlestone and A.~Ovadya.
\newblock The tension between openness and prudence in {AI} research.
\newblock \emph{CoRR}, abs/1910.01170, 2019.
\newblock URL \url{http://arxiv.org/abs/1910.01170}.

\bibitem[Widder et~al.(2022)Widder, Nafus, Dabbish, and
  Herbsleb]{10.1145/3531146.3533779}
D.~G. Widder, D.~Nafus, L.~Dabbish, and J.~Herbsleb.
\newblock Limits and possibilities for “ethical ai” in open source: A study
  of deepfakes.
\newblock In \emph{2022 ACM Conference on Fairness, Accountability, and
  Transparency}, FAccT '22, page 2035–2046, New York, NY, USA, 2022.
  Association for Computing Machinery.
\newblock ISBN 9781450393522.
\newblock \doi{10.1145/3531146.3533779}.
\newblock URL \url{https://doi.org/10.1145/3531146.3533779}.

\bibitem[Workshop et~al.(2022)Workshop, {:}, Scao, Fan, Akiki, Pavlick, Ilić,
  Hesslow, Castagné, Luccioni, Yvon, Gallé, Tow, Rush, Biderman, Webson,
  Ammanamanchi, Wang, Sagot, Muennighoff, del Moral, Ruwase, Bawden, Bekman,
  McMillan-Major, Beltagy, Nguyen, Saulnier, Tan, Suarez, Sanh, Laurençon,
  Jernite, Launay, Mitchell, Raffel, Gokaslan, Simhi, Soroa, Aji, Alfassy,
  Rogers, Nitzav, Xu, Mou, Emezue, Klamm, Leong, van Strien, Adelani, Radev,
  Ponferrada, Levkovizh, Kim, Natan, De~Toni, Dupont, Kruszewski, Pistilli,
  Elsahar, Benyamina, Tran, Yu, Abdulmumin, Johnson, Gonzalez-Dios, de~la Rosa,
  Chim, Dodge, Zhu, Chang, Frohberg, Tobing, Bhattacharjee, Almubarak, Chen,
  Lo, Von~Werra, Weber, Phan, allal, Tanguy, Dey, Muñoz, Masoud, Grandury,
  Šaško, Huang, Coavoux, Singh, Jiang, Vu, Jauhar, Ghaleb, Subramani,
  Kassner, Khamis, Nguyen, Espejel, de~Gibert, Villegas, Henderson, Colombo,
  Amuok, Lhoest, Harliman, Bommasani, López, Ribeiro, Osei, Pyysalo, Nagel,
  Bose, Muhammad, Sharma, Longpre, Nikpoor, Silberberg, Pai, Zink, Torrent,
  Schick, Thrush, Danchev, Nikoulina, Laippala, Lepercq, Prabhu, Alyafeai,
  Talat, Raja, Heinzerling, Si, Taşar, Salesky, Mielke, Lee, Sharma, Santilli,
  Chaffin, Stiegler, Datta, Szczechla, Chhablani, Wang, Pandey, Strobelt,
  Fries, Rozen, Gao, Sutawika, Bari, Al-shaibani, Manica, Nayak, Teehan,
  Albanie, Shen, Ben-David, Bach, Kim, Bers, Fevry, Neeraj, Thakker, Raunak,
  Tang, Yong, Sun, Brody, Uri, Tojarieh, Roberts, Chung, Tae, Phang, Press, Li,
  Narayanan, Bourfoune, Casper, Rasley, Ryabinin, Mishra, Zhang, Shoeybi,
  Peyrounette, Patry, Tazi, Sanseviero, von Platen, Cornette, Lavallée,
  Lacroix, Rajbhandari, Gandhi, Smith, Requena, Patil, Dettmers, Baruwa, Singh,
  Cheveleva, Ligozat, Subramonian, Névéol, Lovering, Garrette, Tunuguntla,
  Reiter, Taktasheva, Voloshina, Bogdanov, Winata, Schoelkopf, Kalo, Novikova,
  Forde, Clive, Kasai, Kawamura, Hazan, Carpuat, Clinciu, Kim, Cheng, Serikov,
  Antverg, van~der Wal, Zhang, Zhang, Gehrmann, Mirkin, Pais, Shavrina,
  Scialom, Yun, Limisiewicz, Rieser, Protasov, Mikhailov, Pruksachatkun,
  Belinkov, Bamberger, Kasner, Rueda, Pestana, Feizpour, Khan, Faranak, Santos,
  Hevia, Unldreaj, Aghagol, Abdollahi, Tammour, HajiHosseini, Behroozi,
  Ajibade, Saxena, Ferrandis, Contractor, Lansky, David, Kiela, Nguyen, Tan,
  Baylor, Ozoani, Mirza, Ononiwu, Rezanejad, Jones, Bhattacharya, Solaiman,
  Sedenko, Nejadgholi, Passmore, Seltzer, Sanz, Dutra, Samagaio, Elbadri,
  Mieskes, Gerchick, Akinlolu, McKenna, Qiu, Ghauri, Burynok, Abrar, Rajani,
  Elkott, Fahmy, Samuel, An, Kromann, Hao, Alizadeh, Shubber, Wang, Roy,
  Viguier, Le, Oyebade, Le, Yang, Nguyen, Kashyap, Palasciano, Callahan,
  Shukla, Miranda-Escalada, Singh, Beilharz, Wang, Brito, Zhou, Jain, Xu,
  Fourrier, Periñán, Molano, Yu, Manjavacas, Barth, Fuhrimann, Altay, Bayrak,
  Burns, Vrabec, Bello, Dash, Kang, Giorgi, Golde, Posada, Sivaraman,
  Bulchandani, Liu, Shinzato, de~Bykhovetz, Takeuchi, Pàmies, Castillo,
  Nezhurina, Sänger, Samwald, Cullan, Weinberg, De~Wolf, Mihaljcic, Liu,
  Freidank, Kang, Seelam, Dahlberg, Broad, Muellner, Fung, Haller,
  Chandrasekhar, Eisenberg, Martin, Canalli, Su, Su, Cahyawijaya, Garda,
  Deshmukh, Mishra, Kiblawi, Ott, Sang-aroonsiri, Kumar, Schweter, Bharati,
  Laud, Gigant, Kainuma, Kusa, Labrak, Bajaj, Venkatraman, Xu, Xu, Xu, Tan,
  Xie, Ye, Bras, Belkada, and Wolf]{https://doi.org/10.48550/arxiv.2211.05100}
B.~Workshop, {:}, T.~L. Scao, A.~Fan, C.~Akiki, E.~Pavlick, S.~Ilić,
  D.~Hesslow, R.~Castagné, A.~S. Luccioni, F.~Yvon, M.~Gallé, J.~Tow, A.~M.
  Rush, S.~Biderman, A.~Webson, P.~S. Ammanamanchi, T.~Wang, B.~Sagot,
  N.~Muennighoff, A.~V. del Moral, O.~Ruwase, R.~Bawden, S.~Bekman,
  A.~McMillan-Major, I.~Beltagy, H.~Nguyen, L.~Saulnier, S.~Tan, P.~O. Suarez,
  V.~Sanh, H.~Laurençon, Y.~Jernite, J.~Launay, M.~Mitchell, C.~Raffel,
  A.~Gokaslan, A.~Simhi, A.~Soroa, A.~F. Aji, A.~Alfassy, A.~Rogers, A.~K.
  Nitzav, C.~Xu, C.~Mou, C.~Emezue, C.~Klamm, C.~Leong, D.~van Strien, D.~I.
  Adelani, D.~Radev, E.~G. Ponferrada, E.~Levkovizh, E.~Kim, E.~B. Natan,
  F.~De~Toni, G.~Dupont, G.~Kruszewski, G.~Pistilli, H.~Elsahar, H.~Benyamina,
  H.~Tran, I.~Yu, I.~Abdulmumin, I.~Johnson, I.~Gonzalez-Dios, J.~de~la Rosa,
  J.~Chim, J.~Dodge, J.~Zhu, J.~Chang, J.~Frohberg, J.~Tobing,
  J.~Bhattacharjee, K.~Almubarak, K.~Chen, K.~Lo, L.~Von~Werra, L.~Weber,
  L.~Phan, L.~B. allal, L.~Tanguy, M.~Dey, M.~R. Muñoz, M.~Masoud,
  M.~Grandury, M.~Šaško, M.~Huang, M.~Coavoux, M.~Singh, M.~T.-J. Jiang,
  M.~C. Vu, M.~A. Jauhar, M.~Ghaleb, N.~Subramani, N.~Kassner, N.~Khamis,
  O.~Nguyen, O.~Espejel, O.~de~Gibert, P.~Villegas, P.~Henderson, P.~Colombo,
  P.~Amuok, Q.~Lhoest, R.~Harliman, R.~Bommasani, R.~L. López, R.~Ribeiro,
  S.~Osei, S.~Pyysalo, S.~Nagel, S.~Bose, S.~H. Muhammad, S.~Sharma,
  S.~Longpre, S.~Nikpoor, S.~Silberberg, S.~Pai, S.~Zink, T.~T. Torrent,
  T.~Schick, T.~Thrush, V.~Danchev, V.~Nikoulina, V.~Laippala, V.~Lepercq,
  V.~Prabhu, Z.~Alyafeai, Z.~Talat, A.~Raja, B.~Heinzerling, C.~Si, D.~E.
  Taşar, E.~Salesky, S.~J. Mielke, W.~Y. Lee, A.~Sharma, A.~Santilli,
  A.~Chaffin, A.~Stiegler, D.~Datta, E.~Szczechla, G.~Chhablani, H.~Wang,
  H.~Pandey, H.~Strobelt, J.~A. Fries, J.~Rozen, L.~Gao, L.~Sutawika, M.~S.
  Bari, M.~S. Al-shaibani, M.~Manica, N.~Nayak, R.~Teehan, S.~Albanie, S.~Shen,
  S.~Ben-David, S.~H. Bach, T.~Kim, T.~Bers, T.~Fevry, T.~Neeraj, U.~Thakker,
  V.~Raunak, X.~Tang, Z.-X. Yong, Z.~Sun, S.~Brody, Y.~Uri, H.~Tojarieh,
  A.~Roberts, H.~W. Chung, J.~Tae, J.~Phang, O.~Press, C.~Li, D.~Narayanan,
  H.~Bourfoune, J.~Casper, J.~Rasley, M.~Ryabinin, M.~Mishra, M.~Zhang,
  M.~Shoeybi, M.~Peyrounette, N.~Patry, N.~Tazi, O.~Sanseviero, P.~von Platen,
  P.~Cornette, P.~F. Lavallée, R.~Lacroix, S.~Rajbhandari, S.~Gandhi,
  S.~Smith, S.~Requena, S.~Patil, T.~Dettmers, A.~Baruwa, A.~Singh,
  A.~Cheveleva, A.-L. Ligozat, A.~Subramonian, A.~Névéol, C.~Lovering,
  D.~Garrette, D.~Tunuguntla, E.~Reiter, E.~Taktasheva, E.~Voloshina,
  E.~Bogdanov, G.~I. Winata, H.~Schoelkopf, J.-C. Kalo, J.~Novikova, J.~Z.
  Forde, J.~Clive, J.~Kasai, K.~Kawamura, L.~Hazan, M.~Carpuat, M.~Clinciu,
  N.~Kim, N.~Cheng, O.~Serikov, O.~Antverg, O.~van~der Wal, R.~Zhang, R.~Zhang,
  S.~Gehrmann, S.~Mirkin, S.~Pais, T.~Shavrina, T.~Scialom, T.~Yun,
  T.~Limisiewicz, V.~Rieser, V.~Protasov, V.~Mikhailov, Y.~Pruksachatkun,
  Y.~Belinkov, Z.~Bamberger, Z.~Kasner, A.~Rueda, A.~Pestana, A.~Feizpour,
  A.~Khan, A.~Faranak, A.~Santos, A.~Hevia, A.~Unldreaj, A.~Aghagol,
  A.~Abdollahi, A.~Tammour, A.~HajiHosseini, B.~Behroozi, B.~Ajibade,
  B.~Saxena, C.~M. Ferrandis, D.~Contractor, D.~Lansky, D.~David, D.~Kiela,
  D.~A. Nguyen, E.~Tan, E.~Baylor, E.~Ozoani, F.~Mirza, F.~Ononiwu,
  H.~Rezanejad, H.~Jones, I.~Bhattacharya, I.~Solaiman, I.~Sedenko,
  I.~Nejadgholi, J.~Passmore, J.~Seltzer, J.~B. Sanz, L.~Dutra, M.~Samagaio,
  M.~Elbadri, M.~Mieskes, M.~Gerchick, M.~Akinlolu, M.~McKenna, M.~Qiu,
  M.~Ghauri, M.~Burynok, N.~Abrar, N.~Rajani, N.~Elkott, N.~Fahmy, O.~Samuel,
  R.~An, R.~Kromann, R.~Hao, S.~Alizadeh, S.~Shubber, S.~Wang, S.~Roy,
  S.~Viguier, T.~Le, T.~Oyebade, T.~Le, Y.~Yang, Z.~Nguyen, A.~R. Kashyap,
  A.~Palasciano, A.~Callahan, A.~Shukla, A.~Miranda-Escalada, A.~Singh,
  B.~Beilharz, B.~Wang, C.~Brito, C.~Zhou, C.~Jain, C.~Xu, C.~Fourrier, D.~L.
  Periñán, D.~Molano, D.~Yu, E.~Manjavacas, F.~Barth, F.~Fuhrimann, G.~Altay,
  G.~Bayrak, G.~Burns, H.~U. Vrabec, I.~Bello, I.~Dash, J.~Kang, J.~Giorgi,
  J.~Golde, J.~D. Posada, K.~R. Sivaraman, L.~Bulchandani, L.~Liu, L.~Shinzato,
  M.~H. de~Bykhovetz, M.~Takeuchi, M.~Pàmies, M.~A. Castillo, M.~Nezhurina,
  M.~Sänger, M.~Samwald, M.~Cullan, M.~Weinberg, M.~De~Wolf, M.~Mihaljcic,
  M.~Liu, M.~Freidank, M.~Kang, N.~Seelam, N.~Dahlberg, N.~M. Broad,
  N.~Muellner, P.~Fung, P.~Haller, R.~Chandrasekhar, R.~Eisenberg, R.~Martin,
  R.~Canalli, R.~Su, R.~Su, S.~Cahyawijaya, S.~Garda, S.~S. Deshmukh,
  S.~Mishra, S.~Kiblawi, S.~Ott, S.~Sang-aroonsiri, S.~Kumar, S.~Schweter,
  S.~Bharati, T.~Laud, T.~Gigant, T.~Kainuma, W.~Kusa, Y.~Labrak, Y.~S. Bajaj,
  Y.~Venkatraman, Y.~Xu, Y.~Xu, Y.~Xu, Z.~Tan, Z.~Xie, Z.~Ye, M.~Bras,
  Y.~Belkada, and T.~Wolf.
\newblock Bloom: A 176b-parameter open-access multilingual language model,
  2022.
\newblock URL \url{https://arxiv.org/abs/2211.05100}.

\bibitem[Xue et~al.(2020)Xue, He, Wang, and
  Liu]{DBLP:journals/corr/abs-2011-13564}
M.~Xue, C.~He, J.~Wang, and W.~Liu.
\newblock {DNN} intellectual property protection: Taxonomy, methods, attack
  resistance, and evaluations.
\newblock \emph{CoRR}, abs/2011.13564, 2020.
\newblock URL \url{https://arxiv.org/abs/2011.13564}.

\bibitem[Zhang et~al.(2018)Zhang, Gu, Jang, Wu, Stoecklin, Huang, and
  Molloy]{10.1145/3196494.3196550}
J.~Zhang, Z.~Gu, J.~Jang, H.~Wu, M.~P. Stoecklin, H.~Huang, and I.~Molloy.
\newblock Protecting intellectual property of deep neural networks with
  watermarking.
\newblock In \emph{Proceedings of the 2018 on Asia Conference on Computer and
  Communications Security}, ASIACCS '18, page 159–172, New York, NY, USA,
  2018. Association for Computing Machinery.
\newblock ISBN 9781450355766.
\newblock \doi{10.1145/3196494.3196550}.
\newblock URL \url{https://doi.org/10.1145/3196494.3196550}.

\bibitem[Zhang et~al.(2019)Zhang, Xu, Cuesta{-}Infante, and
  Veeramachaneni]{DBLP:journals/corr/abs-1909-01285}
K.~A. Zhang, L.~Xu, A.~Cuesta{-}Infante, and K.~Veeramachaneni.
\newblock Robust invisible video watermarking with attention.
\newblock \emph{CoRR}, abs/1909.01285, 2019.
\newblock URL \url{http://arxiv.org/abs/1909.01285}.

\bibitem[Zhang et~al.(2022)Zhang, Roller, Goyal, Artetxe, Chen, Chen, Dewan,
  Diab, Li, Lin, Mihaylov, Ott, Shleifer, Shuster, Simig, Koura, Sridhar, Wang,
  and Zettlemoyer]{https://doi.org/10.48550/arxiv.2205.01068}
S.~Zhang, S.~Roller, N.~Goyal, M.~Artetxe, M.~Chen, S.~Chen, C.~Dewan, M.~Diab,
  X.~Li, X.~V. Lin, T.~Mihaylov, M.~Ott, S.~Shleifer, K.~Shuster, D.~Simig,
  P.~S. Koura, A.~Sridhar, T.~Wang, and L.~Zettlemoyer.
\newblock Opt: Open pre-trained transformer language models, 2022.
\newblock URL \url{https://arxiv.org/abs/2205.01068}.

\end{thebibliography}
\newpage
\appendix

\section{Developer and Logo Key} \label{Appendix:A}
\begin{figure}[h!]
  \centering
  \includegraphics[width=\textwidth, trim={2cm 3cm 2cm 2.5cm},clip]{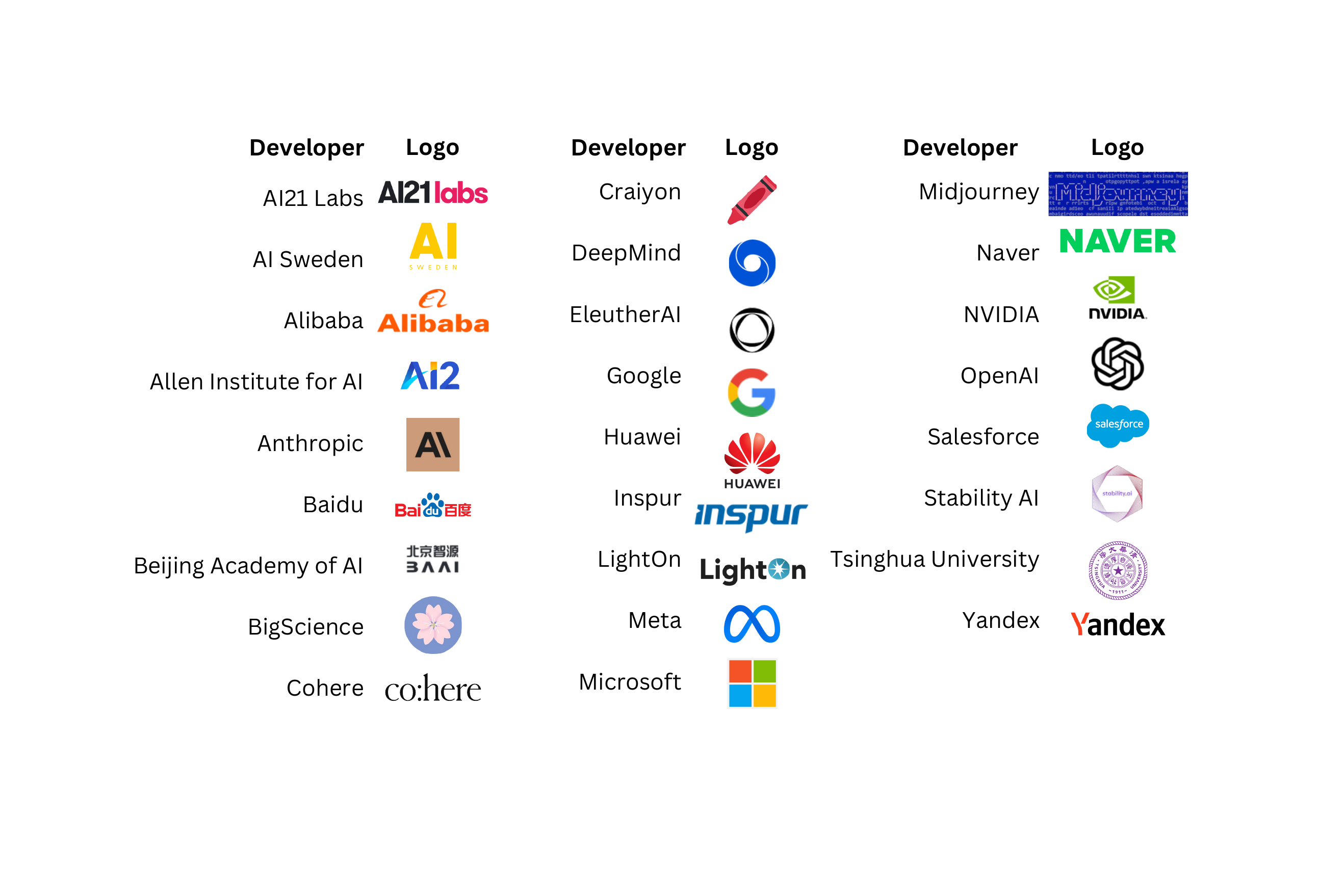}
\end{figure}

\end{document}